\documentclass[journal,twoside,web]{ieeecolor}

\usepackage{generic}
\usepackage{amsmath,amssymb,amsfonts}
\usepackage{algorithmic}
\usepackage{graphicx}
\usepackage{textcomp}
\usepackage{mathrsfs}
\usepackage[dvipsnames]{xcolor}
\usepackage{cases}
\usepackage{mathtools}
\usepackage{caption}
\usepackage{makeidx}
\usepackage{cuted}
\usepackage{float}
\usepackage{hyperref}
\usepackage{dblfloatfix}
\usepackage{setspace}
\usepackage{optidef}
\usepackage{cancel}

\usepackage{amsthm}

\newtheorem{thm}{Theorem}[section]
\newtheorem{lem}[thm]{Lemma}
\newtheorem{prop}[thm]{Proposition}
\newtheorem{cor}{Corollary}[section]
\newtheorem{conj}{Conjecture}[section]
\theoremstyle{definition}
\newtheorem{defn}{Definition}[section]
\newtheorem{exmp}{Example}[section]
\newtheorem{rem}{Remark}
\newtheorem{assumption}{Assumption} 

\newtheorem*{prop*}{Proposition}
\newtheorem*{thm*}{Theorem}

\newcommand{\nd}[1]{\mathsf{#1}}

\def\BibTeX{{\rm B\kern-.05em{\sc i\kern-.025em b}\kern-.08em
    T\kern-.1667em\lower.7ex\hbox{E}\kern-.125emX}}
\begin{document}
\title{Spiking control systems for soft robotics: \\
a rhythmic case study in a soft robotic crawler}
\author{Juncal Arbelaiz, \IEEEmembership{Member, IEEE}, Alessio Franci, \IEEEmembership{Member, IEEE}, Naomi Ehrich Leonard, \IEEEmembership{Fellow, IEEE}, Rodolphe Sepulchre, \IEEEmembership{Fellow, IEEE}, Bassam Bamieh, \IEEEmembership{Fellow, IEEE}
\thanks{
J. Arbelaiz acknowledges the support of the Schmidt Science Fellowship, in
partnership with the Rhodes Trust.}
\thanks{
J. Arbelaiz is at the Center for Statistics and Machine Learning at Princeton University,  NJ,  08542,  USA (e-mail: \texttt{jarbelaiz@schmidtsciencefellows.org}). }
\thanks{A. Franci is with the Department of Electrical Engineering and Computer Science, University of Liege, and WEL Research Institute, Wavre, Belgium (e-mail: \texttt{afranci@uliege.be})}
\thanks{N. E. Leonard is with the Department of Mechanical and Aerospace Engineering, Princeton University, NJ, USA (e-mail: \texttt{naomi@princeton.edu})}
\thanks{R. Sepulchre is with the Department of Electrical Engineering, KU Leuven, B-3001 Leuven, Belgium (e-mail: \texttt{rodolphe.sepulchre@kuleuven.be})}
\thanks{B.  Bamieh is with the department of Mechanical Engineering at the University of California at Santa Barbara,  CA,  93106-2014 USA (e-mail: \texttt{bamieh@ucsb.edu}).}
\thanks{}
}

\maketitle

\begin{abstract}
Inspired by spiking neural feedback, we propose a spiking controller to engineer the locomotion of a soft robotic crawler. Its bistability, akin to neural fast positive feedback, combined with a sensorimotor slow negative feedback loop, generates rhythmic spiking and self-sustained peristaltic locomotion. 
We analytically characterize the
local equilibrium bifurcations induced by the sensorimotor gain and
complement this analysis through numerical continuation of periodic
orbits. The resulting bifurcation diagram reveals 
the emergence of qualitatively distinct solutions, including resting and crawling behaviors. For a representative regime with separated mechanical and electrical timescales, Geometric Singular Perturbation Theory reveals the geometry of the relaxation oscillations leading to endogenous crawling. 
Within this singularly perturbed regime, we formulate and analytically solve an optimization problem, proving that locomotion speed is maximized at mechanical resonance through a
matching of neuromechanical scales. Given the importance and ubiquity of rhythms and waves in soft-bodied locomotion, we envision that spiking control systems could be utilized in a variety of soft-robotic morphologies and modular distributed architectures. 
\end{abstract}

\begin{IEEEkeywords}
Spiking control systems, soft robotics, biologically-inspired methods
\end{IEEEkeywords}

\section{Introduction}
\label{sec:introduction}

 The elasticity of the musculoskeletal system is fundamental to effective and efficient animal performance 
\cite{DellaSantina2023,Roberts2016}. Inspired by this observation, 
soft robotics seeks to leverage body compliance to efficiently handle diverse environments and uncertainty in a safer, more energy-efficient, and robust manner than with state-of-the-art rigid robotic technology \cite{Hauser2023}.  
Soft robotics promises exciting new capabilities, offering a frontier in robotics with enormous potential
for minimally invasive and laparoscopic surgeries \cite{Cianchetti2018}, inspection tasks in unstructured environments, and grasping of delicate objects \cite{Laschi2016}.
While the promise is large, the technology is still in its infancy as the variability in robot morphology and scale, the challenge of accurately modeling their complex dynamics, and the limited onboard power and computation make the design and implementation of feedback control architectures a daunting task \cite{DellaSantina2023}.  
 Control theory tailored to soft robotics is an important step towards unlocking its potential.

The control of rhythms across scales is central to biological and robotic function \cite{Sepulchre2019,Che2021}.  
Rhythmic movements span a large portion of the animal and robotic locomotive repertoire \cite{Simoni2007}.  The generation of such rhythms---typically attributed to Central Pattern Generators  \cite{Ijspeert2008}, \cite[Ch. 3]{GolubitskyM.andStewart2002}---
requires coordinated control signals mediating the interactions of the musculoskeletal and nervous systems with the environment. 
While gait design has historically focused on rigid robotics \cite{DellaSantina2023}, 
a principled framework in the emerging field of soft robotics is lagging behind: controllers for soft robotic gait sequences are often hand-crafted---a non-trivial and time-consuming process \cite{Ketchum2023}---or designed in open-loop \cite{Shen2024,Gusty2025}.

Among the different strategies used by soft-bodied animals and robots for rhythmic locomotion, crawling is a common one. Earthworms in nature \cite{Tanaka2012} and robotic crawlers in engineering (see, e.g., 
\cite{Ze2022,Seok2013,Earnhardt2025}) typically move by propagating peristaltic waves along their bodies, 
combined with anisotropic friction with the substrate. Compliant crawling robots are particularly relevant for inspection and navigation in confined, cluttered, or hazardous environments, such as pipes and ducts, debris in search-and-rescue operations, or spaces unsafe for conventional rigid robots. In this work, we propose and analyze the use of an excitable feedback controller \cite{Sepulchre2019,Sepulchre2018,Sepulchre2022} in a soft robotic crawler to engineer its peristaltic locomotion. Our {\it spiking}  controller is inspired by the FitzHugh-Nagumo model of neural excitability \cite{Sepulchre2018}. Related recent work \cite{Motta2025} uses a piecewise-linear FitzHugh-Nagumo model to control passive actuators and coordinate robotic-hand locomotion.

Spiking control systems \cite{Sepulchre2022} combine attractive properties of analog and digital controllers: they offer robustness to model uncertainty and component degradation, they can be modulated for control across scales, and they possess the continuous adaptation capability of analog systems with the discrete reliability of digital automata, all with potentially reduced power consumption. Such features are well suited to the inherent challenges of synthesis and implementation of feedback control for soft robots. While some neuro-inspired and spiking controllers for soft crawlers have been proposed (see, e.g., \cite{Zhang,Paoletti2014} 
 and \cite{Asawalertsak2023}), their evaluation typically relies on experiments, with little to no control-theoretic guarantees. A principled framework grounded in rigorous control-theoretic tools remains limited for \textit{embodied spiking systems}. We start bridging this gap with the current work, where
we mathematically analyze a single-input single-output spiking controller for a two-segmented spiking soft robotic crawler. We characterize its transition from resting to a locomotive regime, highlighting its tunability by parameters in the excitable crawler dynamics, and providing principles for parameter selection and  optimal operation. 
The two-segmented excitable crawler exhibits rich dynamics and constitutes a fundamental building block to scale towards more complex distributed multi-segmented soft robotic architectures. Extensions to modular multi-segmented crawlers will require the synthesis of multi-input multi-output spiking controllers, which can rely on distributed control theory \cite{Bamieh,Arbelaiz2022,Arbelaiz2025, Arbelaiz2021} and the theory of fast and flexible multi-agent multi-option decision making \cite{Leonard2024, Xul}.
Since coordinated body contractions and expansions are a common mechanism for locomotion in a variety of soft-bodied robotic morphologies (see, e.g.,
\cite{Yang2021}), we envision that the control paradigm introduced in this work could extend beyond the specific crawler morphology considered here.

\paragraph*{Main Contributions} This paper substantially extends \cite{Arbelaiz_2024}, which introduced the spiking controller for a soft crawler and provided preliminary evidence of self-sustained locomotion. We characterize the organizing bifurcations of the \textit{excitable crawler}—the closed-loop system formed by an excitable feedback controller and soft robot—as the sensorimotor gain varies, providing guidelines for operating-point selection. We then identify a \textit{regime with separated electrical and mechanical timescales} and use Geometric Singular Perturbation analysis to characterize the geometry of the self-sustained peristaltic relaxation oscillations. Finally, using describing function analysis, we formulate and solve an optimization problem for the crawler’s average center-of-mass speed, showing that \textit{optimal crawling occurs at mechanical resonance} through a \textit{neuromechanical matching} condition. At this optimum, the fundamental harmonics of strain rate and actuation force are phase-aligned within the harmonic approximation,
making the instantaneous actuator power non-negative and maximizing its average over a crawl. Taken together, these results show how proprioceptive spiking feedback can endogenously generate a stable locomotion rhythm, provide a low-dimensional handle for gait modulation through the sensorimotor gain, and match actuation timing to compliant-body mechanics.

\paragraph*{Paper Structure}§II summarizes mathematical preliminaries. §III introduces and nondimensionalizes the excitable crawler dynamics. §IV analyzes organizing bifurcations underlying resting and crawling behaviors, with proofs in Appendix I. §V develops the slow–fast and Geometric Singular Perturbation analysis, with proofs in Appendix II. §VI uses describing-function and harmonic-balance analysis (Appendix III) to derive optimal controller parameters and establish resonance optimality. §VII concludes.

\section{Mathematical Preliminaries}
\label{sec:Mathematical_pre}

\paragraph{Notation} 

Lowercase regular (e.g., $x$), lowercase bold (e.g., $\boldsymbol{x}$), and capital fonts (e.g., $A$) denote scalars, vectors, and matrices, respectively.  $\mathsf{Sans \; serif}$ font denotes dimensionless variables, e.g., if $\boldsymbol{x}$ is a dimensional vector, $\boldsymbol{\nd{x}}$ is its dimensionless counterpart. Dimensionless groups are denoted by $\pi_{\nd{i}}$, unless otherwise noted. 
 We use $\lambda$ for eigenvalues and $\nabla_{\boldsymbol{x}} \boldsymbol{f}(\boldsymbol{x}_0; \mu)$ for the Jacobian of  $\boldsymbol{f}: \mathbb{R}^n \to \mathbb{R}^n$ at $\boldsymbol{x}_0$ and parameterized by $\mu$.  $\mathbb{R}_+$ ($\mathbb{R}_{++}$) denotes the nonnegative (positive) reals. 
 $\mathbb{I}_n$ is the identity matrix of dimension $n$. $C^\infty(U)$ is the class of infinitely differentiable functions on $U \subseteq \mathbb{R}^n$.  
 The imaginary unit is denoted by $\mathbf{i}$. 
 For $\alpha \in \mathbb{C}$, $\Re(\alpha)$ and $\Im(\alpha)$ denote its real and imaginary parts, respectively, and $\bar{\alpha}$ its complex conjugate. For $\boldsymbol{v}_1, \boldsymbol{v}_2 \in \mathbb{C}^n$, $\langle \boldsymbol{v}_1, \boldsymbol{v}_2 \rangle := \bar{\boldsymbol{v}}_1^\top \boldsymbol{v}_2$. 
 
\paragraph{Groups, equivariances, and symmetries} 
Consider ODEs
\begin{equation}
    \dot{\boldsymbol{x}}(t) = \boldsymbol{f}(\boldsymbol{x}(t); \boldsymbol{\mu}), \; \boldsymbol{f}: \mathbb{R}^n \times \mathbb{R}^r \to \mathbb{R}^n, \boldsymbol{f} \in C^{\infty},
    \label{eq:general_dynamics}
\end{equation}
where $\boldsymbol{x}(t) \in \mathbb{R}^n$ denotes the state and $\boldsymbol{\mu} \in \mathbb{R}^r$ are parameters.
\begin{defn}[Symmetry, from \cite{GolubitskyM.andStewart2002}]
\label{defn:equivariant_mapping}
    The group element $g \in \mathbb{G}$
    is a symmetry of \eqref{eq:general_dynamics} if for every solution $\boldsymbol{x}$ of \eqref{eq:general_dynamics}, $g \boldsymbol{x}$ is also a solution.
\end{defn}

Equivalently, since solutions exist for arbitrary initial conditions (Cauchy–Lipschitz theorem):
$
 \boldsymbol{f}\big(g \boldsymbol{x} \big) = g \boldsymbol{f}\big( \boldsymbol{x} \big) \, \forall \, \text{solutions } \boldsymbol{x} \Leftrightarrow 
 g \boldsymbol{f}\big( \boldsymbol{x}(t)\big) =  \boldsymbol{f}\big( g \boldsymbol{x}(t)\big) \,\forall \boldsymbol{x}(t) \in \mathbb{R}^n,
 $
and we say that $\boldsymbol{f}$ \textit{commutes with $g$} or \textit{$\boldsymbol{f}$ is $g$-equivariant}.

\begin{defn}[$\mathbb{G}$-equivariant mapping, from \cite{GolubitskyM.andStewart2002}]
\label{defn:equivariant_mapping_2}
    Let the group $\mathbb{G}$ act on $\mathbb{R}^n$ and let $\boldsymbol{f}: \mathbb{R}^n \to \mathbb{R}^n$. Then, $\boldsymbol{f}$ is $\mathbb{G}$-equivariant if $g \boldsymbol{f}\big(  \boldsymbol{x}\big) = \boldsymbol{f}\big( g \boldsymbol{x}\big), \forall g \in \mathbb{G}, \; \forall \boldsymbol{x} \in \mathbb{R}^n.$
\end{defn}

\paragraph{Equilibria, bifurcations, and related considerations}  
For the system \eqref{eq:general_dynamics}, an \textit{equilibrium} or \textit{fixed point}  $\boldsymbol{x}_* \in \mathbb{R}^n$ satisfies $\boldsymbol{f}(\boldsymbol{x}_*; \boldsymbol{\mu}) = \mathbf{0}$.  If all the eigenvalues of  
$\nabla_{\boldsymbol{x}} \boldsymbol{f}(\boldsymbol{x}_*; \boldsymbol{\mu})$  have non-zero real parts, $\boldsymbol{x}_*$ is \textit{hyperbolic} and the stability of $\boldsymbol{x}_*$ can be assessed by analyzing the linearization of \eqref{eq:general_dynamics} at $\boldsymbol{x}_*$ (by Hartman's theorem, \cite[Thm 1.3.1]{Guckenheimer1983}).
If $\boldsymbol{x}_*$ is \textit{non-hyperbolic}, 
it is necessary to consider nonlinear terms for its stability assessment. In 
\eqref{eq:general_dynamics} such non-hyperbolicities might arise at particular parameter values, $\boldsymbol{\mu}_*$, that change topological features of the flow. We call $\boldsymbol{\mu}_*$ the \textit{bifurcation value} and such parametrized topological changes in the flow are known as \textit{bifurcations}. The pair $(\boldsymbol{x}_*; \boldsymbol{\mu}_*)$ is referred to as a \textit{bifurcation point} and the parameter that is varied is the \textit{bifurcation parameter} \cite[Ch. 3]{Guckenheimer1983}.  A necessary condition for a local bifurcation of an equilibrium is that the equilibrium lose hyperbolicity.
 If \eqref{eq:general_dynamics} has a bifurcation point, the nontrivial dynamics near the corresponding fixed point happen within the \textit{center manifold}\footnote{The center manifold need not be unique. See \cite[Thm 3.2.1]{Guckenheimer1983}.}.

When the dimension of the center manifold is smaller than that of the full system, bifurcation classification often relies on a local reduction to the critical directions, for example through the \textit{center manifold reduction} \cite[\S 3.2]{Guckenheimer1983} or \textit{Lyapunov-Schmidt reduction} \cite[Ch. 1, 7, 8]{Guckenheimer1983}, often applied together with a local series expansion approximating the reduced equations close to the bifurcation point. 
 The procedure to classify a bifurcation and its criticality often involves projections of directional derivatives of the vector field $\boldsymbol{f}$ at the bifurcation point into the left null eigenspace of $\nabla_{\boldsymbol{x}} \boldsymbol{f}(\boldsymbol{x}_*; \boldsymbol{\mu}_*)$, and the recognition of the corresponding \textit{normal form} of the bifurcation \cite{Guckenheimer1983, Golubitsky1985}. We introduce next the definition of \textit{directional derivative of a vector field}.

\begin{defn}[Directional derivatives of a vector field]
\label{def:directional_derivative}
    Consider a parametrized vector field $\boldsymbol{f}(\boldsymbol{x}; \boldsymbol{\mu}): \mathbb{R}^n \times \mathbb{R}^r \to \mathbb{R}^n$, sufficiently smooth. 
    Given a tuple of vectors 
    $\mathcal{V} =(\boldsymbol{v}_1,\dots,\boldsymbol{v}_k) \in (\mathbb{R}^n)^k$, the $k-$th order directional derivative of $\boldsymbol{f}$ along $\mathcal{V}$ at $(\boldsymbol{x}_0; \boldsymbol{\mu}_0)$ is
    \begin{small}
    $
    \big( \mathrm{d}^k \boldsymbol{f} \big)_{(\boldsymbol{x}_0; \boldsymbol{\mu}_0)}  [\boldsymbol{v}_1, \boldsymbol{v}_2, \dots, \boldsymbol{v}_k] 
     = 
         \frac{\partial}{\partial t_1} \frac{\partial}{\partial t_2} \dots \frac{\partial}{\partial t_k} \boldsymbol{f}\Big( \boldsymbol{x}_0 + \sum_{i = 1}^k t_i \boldsymbol{v}_i ; \boldsymbol{\mu}_0 \Big)\Big\vert_{t_1 = \dots = t_k = 0} 
      = \sum_{i_1, \dots, i_k = 1}^n \frac{\partial^k \boldsymbol{f}}{\partial x_{i_1} \dots \partial x_{i_k}}(\boldsymbol{x}_0; \boldsymbol{\mu}_0)(\boldsymbol{v}_1)_{i_1} \dots (\boldsymbol{v}_k)_{i_k}. 
    \label{eq:direc_derivative_multi}
    $
    \end{small}
\end{defn}
The following theorems characterize two elementary bifurcations that will be useful in our work.

\begin{thm}[Pitchfork bifurcation, from \cite{Golubitsky1985}]
\label{thm:pitchfork_bifurcation}
    Suppose that the sufficiently smooth system $\dot{\boldsymbol{x}} = \boldsymbol{f}(\boldsymbol{x}; \mu) $ with $\boldsymbol{x} \in \mathbb{R}^n$ and $\mu \in \mathbb{R}$ satisfies (1) Equilibrium: $\boldsymbol{f}(\boldsymbol{x}_*; \mu_*) = \mathbf{0}$; (2) Simple zero eigenvalue: $J := \nabla_{\boldsymbol{x}} \boldsymbol{f}(\boldsymbol{x}_*; \mu_*) $ has an eigenvalue $\lambda = 0$ with right-eigenvector $\boldsymbol{v}\in \mathbb{R}^n$ 
    and left-eigenvector $\boldsymbol{w} \in \mathbb{R}^n$, 
    normalized to $\langle \boldsymbol{w}, \boldsymbol{v}\rangle = 1$; and (3) Hyperbolicity of the remaining spectrum: the remaining eigenvalues of $J$ have non-zero real parts. Let $P:= \mathbb{I} - \boldsymbol{v}\boldsymbol{w}^\top$ and suppose the following non-degeneracy conditions hold:
  (4) vanishing constant term: $\langle \boldsymbol{w},  \partial_{\mu} \boldsymbol{f}(\boldsymbol{x}_*; \mu_*) \rangle = 0 \label{eq:pitchfork_1}$; (5) Nondegenerate parameter-amplitude coupling:
            $\langle\boldsymbol{w}, \mathrm{d}\big( \partial_{\mu} \boldsymbol{f}\big)_{(\boldsymbol{x}_*; \mu_*)}\big[\boldsymbol{v}\big]  - (\mathrm{d}^2 \boldsymbol{f})_{(\boldsymbol{x}_*; \mu_*)} [\boldsymbol{v}, J^{-1} P (\partial_\mu \boldsymbol{f})_{(\boldsymbol{x}_*; \mu_*)}] \rangle  \neq 0, \label{eq:pitchfork_2}$ where $J^{-1}$ denotes the inverse of the restriction of $J$ to $\mathrm{range}(P)$; (6)
              Vanishing quadratic term: $\langle \boldsymbol{w}, \mathrm{d}^2 \boldsymbol{f}_
             {(\boldsymbol{x}_*; \mu_*)}\big[\boldsymbol{v},\boldsymbol{v}\big] \rangle  = 0 \label{eq:pitchfork_3}$; and (7) Non-degenerate cubic term:$
               \langle \boldsymbol{w}, \mathrm{d}^3 \boldsymbol{f}_
               {(\boldsymbol{x}_*; \mu_*)}\big[\boldsymbol{v},\boldsymbol{v},\boldsymbol{v}\big] - 3(\mathrm{d}^2 \boldsymbol{f})_{(\boldsymbol{x}_*; \mu_*)} \big[\boldsymbol{v}, J^{-1} P (\mathrm{d}^2 \boldsymbol{f})_{(\boldsymbol{x}_*; \mu_*)} [\boldsymbol{v},\boldsymbol{v}]\big] \rangle  \neq 0 \label{eq:pitchfork_4}$.
    Then, $\dot{\boldsymbol{x}} = \boldsymbol{f}(\boldsymbol{x}; \mu)$ undergoes a pitchfork bifurcation at $\boldsymbol{x}_*$ as the parameter $\mu$ varies through the bifurcation value $\mu = \mu_*$. 
    The criticality of the pitchfork and the stability of the bifurcating branches in the critical direction are determined by the signs of the inner products in (5) and (7). 
\end{thm}

In the absence of symmetry, conditions \textup{(4)} and \textup{(6)} in Theorem \ref{thm:pitchfork_bifurcation} are codimension two, and hence nongeneric in a one-parameter family.  
If the system is $\mathbb{Z}_2$--equivariant, $\boldsymbol{x}_*$ is fixed by the symmetry, and the critical eigendirection is odd under the symmetry, then the reduced bifurcation equation is odd in the amplitude.  
Thus, \textup{(4)} and \textup{(6)} are automatically satisfied, and a pitchfork bifurcation occurs generically.


\begin{thm}[Hopf bifurcation theorem,  Theorem 3.4.2 in \cite{Guckenheimer1983}]
\label{thm:Hopf_bifurcation}
Consider $\dot{\boldsymbol{x}}=\boldsymbol{f}(\boldsymbol{x};\mu)$ with $\boldsymbol{x}\in\mathbb{R}^n$ and $\mu\in\mathbb{R}$, where $\boldsymbol{f}$ is sufficiently smooth.
Assume that $(\boldsymbol{x}_*; \mu_*)$ is an equilibrium, i.e.\ $\boldsymbol{f}(\boldsymbol{x}_*;\mu_*)=\boldsymbol{0}$, and that:
(1) The Jacobian $\nabla_{\boldsymbol{x}}\boldsymbol{f}(\boldsymbol{x}_*;\mu_*)$ has a simple pair of purely imaginary eigenvalues $\pm \mathbf{i}\omega_*$ with $\omega_*>0$; (2) All other eigenvalues have nonzero real part; (3) Along the local equilibrium branch $\boldsymbol{x}(\mu)$ with $\boldsymbol{x}(\mu_*)=\boldsymbol{x}_*$, the corresponding eigenvalues $\lambda(\mu)=\alpha(\mu)+\mathbf{i}\omega(\mu)$ and $\bar{\lambda}(\mu)=\alpha(\mu)-\mathbf{i}\omega(\mu)$ satisfy $d:=\frac{d\alpha}{d\mu}\vert_{\mu=\mu_*}\neq 0
$ (transversality condition).
Then there exists a three-dimensional center manifold through $(\boldsymbol{x}_*,\mu_*)$ in the extended space $\mathbb{R}^n\times\mathbb{R}$.
Moreover, there exist smooth coordinates on this center manifold in which, to cubic order, the reduced dynamics can be written in polar form as
$
\dot r = \big(d(\mu-\mu_*) + a r^2\big)r + \text{h.o.t.},\;
\dot\theta = \omega_* + c(\mu-\mu_*) + b r^2 + \text{h.o.t.}.
$
If $a\neq 0$ (nondegeneracy condition), then a smooth family of small-amplitude periodic orbits bifurcates from $\boldsymbol{x}_*$ as $\mu$ passes through $\mu_*$. 
Furthermore, if $a<0$ (resp.\ $a>0$), the Hopf bifurcation is supercritical (resp.\ subcritical), producing stable (resp.\ unstable) small-amplitude periodic orbits on the parameter side where $d(\mu-\mu_*)>0$ (resp.\ $d(\mu-\mu_*)<0$).
\end{thm}

The sign of the coefficient $a$ of the cubic term in the radial Hopf normal form determines the criticality of the bifurcation. The closely related quantity known as the \textit{first Lyapunov coefficient}, $l_1$, is often used to characterize this cubic nonlinearity---it has the same sign as $a$. Different normalizations of $l_1$ appear in the literature \cite[\S 3]{Kuehn2010}; while their numerical values may differ, their sign is invariant under these normalizations. An explicit formula for the first Lyapunov coefficient is in \cite[eq. (5.34)]{Kuznetsov2023}.

\paragraph{The describing function method}
Describing function analysis---an extended version of the frequency response technique for linear systems---is a quasi-linearization method useful for obtaining an approximation to the response of some nonlinear systems  \cite{Gelb1968}, \cite[Ch. 5]{Slotine1991}. 
The core idea underlying the method is to approximate the periodic output of a nonlinear element by its DC component and \textit{fundamental harmonic}, neglecting higher-order harmonics under the assumption that they are sufficiently attenuated by the filtering properties of the linear components of the system—this is known as the \textit{filtering hypothesis}. Given an input $x(t)  = A \sin (\omega t)$ to a nonlinearity, its periodic output $w(t)$ can be represented in a Fourier series 
\begin{small}
$
    w(t)  = \frac{a_0}{2} + \sum_{n=1}^\infty \Big( a_n \cos(n \omega t) + b_n \sin(n \omega t) \Big)
$
\end{small}
where the coefficients are
\begin{small}
$
    a_n := \frac{1}{\pi} \int_{-\pi}^\pi w(t) \cos(n \omega t)d(\omega t) 
    $
\end{small}
 and 
 \begin{small}
$
    b_n  := \frac{1}{\pi} \int_{-\pi}^\pi w(t) \sin(n \omega t)d(\omega t). 
$
\end{small}
Under the filtering hypothesis, the output $w(t)$ is approximated by its DC component and fundamental harmonic
\begin{small}
$
    w(t)  \approx \frac{a_0}{2} + a_1 \cos(\omega t) + b_1 \sin (\omega t). 
$
\end{small}

\section{The excitable crawler:
Dynamics \& Control}
\label{sec:dynamics_and_control}

We introduce the two elements that compose the \textit{excitable crawler}: the soft crawler's body dynamics and the feedback controller inspired by the FitzHugh-Nagumo model of neural excitability. For convenience in the subsequent analysis, a change of coordinates and the non-dimensionalization of the model are presented as well. 

\subsection{Dynamics of the crawler's body}
\begin{figure}[h]
    \centering
    \includegraphics[width=.49\textwidth]{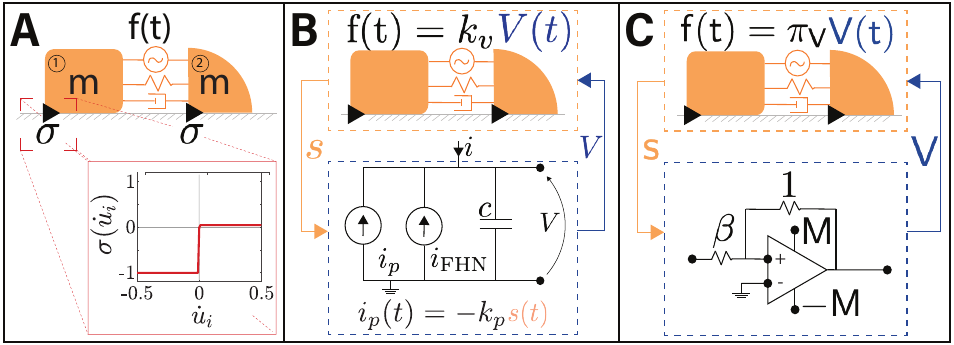}
    \caption{ \begin{small}
    \textbf{(A)} Schematic of the body of the two-segmented soft crawler analyzed in this work. The index 1 (resp., 2) refers to the crawler's tail (resp., head). The 
    inset represents the 
    nonlinear anisotropic friction model as given in \eqref{eq:sigma_def}. \textbf{(B}) The excitable crawler: information flow in the closed-loop system between the crawler's body and the neuromorphic excitable controller proposed in \S\ref{subsec:excitable_controller}. The excitable controller has the
    realization of a flip-flop 
    electrical circuit.  
    \textbf{(C)} The static hysteresis I/O map relating strain and voltage utilized in \S\ref{sec:setting_the_pace} is hardware-realizable with a Schmitt trigger---a comparator with hysteresis. $\beta$ denotes the (dimensionless) switching strain threshold and $ \mathsf{M}$ the (dimensionless) output voltage magnitude. 
    \end{small}
    }
    \label{fig:crawler_schematics}
\end{figure}

 We analyze a two-segmented soft crawler, whose body has a natural length $\ell$---Fig. \ref{fig:crawler_schematics}\textbf{A} provides a schematic. The soft crawler moves in a one-dimensional environment and its body dynamics are modeled as \cite{Che2021,Shen2024, Paoletti2014, Arbelaiz_2024}
\begin{subequations}
\begin{align}
  m  \ddot u_1 &= k (u_2-u_1 ) + b (\dot u_2 - \dot u_1)  - A_\sigma \sigma(\dot u_1 )-f,
  \label{eq:crawler_dynamics_1}\\
  m  \ddot u_2 &= k (u_1-u_2) + b (\dot u_1 - \dot u_2)  - A_\sigma \sigma(\dot u_2) + f,
   \label{eq:crawler_dynamics_2}
\end{align}
where $u_i$ denotes the displacement of mass $i$ ($i = 1,2$) with respect to the natural rest configuration. The time variable, implicit in \eqref{eq:dimensional_crawler_dynamics}, is denoted by $t$ and we use $\dot{u}_i$ to denote $\mathrm{d} u_i/\mathrm{d} t$. The viscoelasticity of the crawler's body is captured through the elastic constant $k$ and the viscous damping constant $b$. $f$ is the actuator signal and $A_\sigma \sigma$ is the frictional force arising due to the interaction of the crawler with the substrate.  $A_\sigma$ is a positive scaling constant and 
\begin{small}
\begin{equation}
    \sigma(\dot{u}):= \frac{\tanh{( \dot{u}/\epsilon_f + \nd{n_f}) - \tanh{(\nd{n_f})}}}{1 + \tanh{(\nd{n_f})}}
    \label{eq:sigma_def}
\end{equation}
\end{small}
\label{eq:dimensional_crawler_dynamics}
\end{subequations}
is a nonlinear anisotropic function of the local speed. Note that $\sigma(0) = 0$ and that $\sup_{\dot{u} \in \mathbb{R}} |\sigma(\dot{u})| = 1$. The dimensionless parameter $\nd{n_f}$ tunes the anisotropy of the frictional force and the dimensional parameter $\epsilon_f$ its slope. Friction anisotropy introduces a preferential direction of motion for the crawler, as forward motion experiences less friction than backward motion. The frictional force model as a function of the local speed $\dot{u}_i$ is depicted in Fig. \ref{fig:crawler_schematics}\textbf{A}. For convenience\footnote{
 In the transformed coordinates, we will avoid working with absolute displacements, leveraging the translation invariance in $u_{\text{com}}$ and focusing on the relative motion with respect to a \textit{moving frame} attached to the crawler's center of mass. In the relative motion resulting after subtraction of the global translation of the center of mass, state entries are  periodic during steady crawling, which facilitates the analysis of the crawling gait as limit cycles naturally live in these relative coordinates.}, we introduce the following change of coordinates:
$
    \begin{bmatrix}
       s \; 
       u_{\text{com}}
    \end{bmatrix}^\top = 
    \mathcal{M}
    \begin{bmatrix}
        u_1\;
        u_2
    \end{bmatrix}^\top 
    \text{ with }
    \mathcal{M}:= 
    \begin{small}
    \begin{bmatrix}
        -1 & 1\\
        \frac{1}{2} &  \frac{1}{2}
    \end{bmatrix},
    \end{small}
$
such that the crawler's segment displacements are recovered as $\textbf{u} = \mathcal{M}^{-1} [s \; u_{\text{com}}]^\top$. $s:=u_2 - u_1$  denotes the \textit{strain} in the crawler's body and  $u_{\text{com}}:= (u_1 + u_2)/2$ the \textit{displacement of its center of mass}. Applying the change of coordinates in the dynamics \eqref{eq:dimensional_crawler_dynamics} yields
\begin{small}
\begin{subequations}
\begin{align}
   m  \ddot{u}_{\text{com}} & =  - \frac{A_\sigma}{2} \bigg( \sigma\Big(\dot{u}_{\text{com}} - \frac{\dot{s}}{2}\Big) + \sigma\Big( \dot{u}_{\text{com}} + \frac{\dot{s}}{2} \Big) \bigg), \label{eq:dimensional_ucom_dynamics}\\
   m\ddot{s} & =  A_\sigma \bigg(\sigma \Big( \dot{u}_{\text{com}} - \frac{\dot{s}}{2} \Big) -  \sigma\Big(\dot{u}_{\text{com}} + \frac{\dot{s}}{2}\Big)\bigg)  -2 (k s + b \dot{s}  - f). \label{eq:dimensional_s_dynamics}
\end{align}
\label{eq:dimensional_electromechanical_dynamics}
\end{subequations}
\end{small}
\eqref{eq:dimensional_ucom_dynamics} shows that the center of mass is only driven by external forces. The \textit{natural frequency}\footnote{The $\omega_n$ defined in \eqref{eq:natural_freq} is the natural frequency of two segments with linear dynamics and respective mass $m$ interconnected by an (undamped) linear spring with elastic constant $k$.} of the crawler's body is
\begin{subequations}
\begin{equation}
    \omega_n := \sqrt{\frac{2k}{m}}.
    \label{eq:natural_freq}
\end{equation}
The \textit{characteristic mechanical scales} in the system are
\begin{equation}
   l_* = \ell,  \; m_* = 2\, m, \text{ and } t_* = 1/\omega_n.
   \label{eq:mechanical_scales}
\end{equation}
\end{subequations}
\vspace{-0.4cm}
\subsection{A neuromorphic excitable controller}
\label{subsec:excitable_controller}

 The objective of the actuator input $f$ in \eqref{eq:dimensional_crawler_dynamics} is the \textit{endogenous} generation of a crawling gait, avoiding exogenous periodic actuator commands. 
 We let the actuation $f$ use \textit{proprioceptive signals} and be proportional to the output voltage $V$ of an electrical circuit
$
    f = k_v \, V,
$
where $k_v \in \mathbb{R}_{++}$ denotes the voltage gain. We design the electrical circuit---depicted in Fig. \ref{fig:crawler_schematics}\textbf{B}---to contain two features  whose intertwined action enables endogenous crawling.  The first is \textit{bistability}, which is achieved by introducing a
     voltage-controlled current source ($i_{\text{FHN}}$) with a localized region of negative conductance inspired by the FitzHugh-Nagumo model of neuronal excitability  \cite{Sepulchre2018}:
     $
         i_{\text{FHN}} = - \kappa \, V^3 + \rho \, V,
         \label{eq:i_FHN}
     $
     where the parameters $\kappa \in \mathbb{R}_{++}$ and $\rho \in \mathbb{R}_{++}$ tune the strength of the global negative feedback and local positive feedback provided by the current source, respectively. Since $f = k_v \, V$ with $k_v \in \mathbb{R}_{++}$, positive (negative) voltage leads to crawler extension (contraction). The second is a
    \textit{ proprioceptive feedback current} ($i_p$), negatively proportional to crawler's strain: 
     $
         i_p  
         = -k_p \, s,
         \label{eq:i_p}
     $
     where $k_p \in \mathbb{R}_{++}$ is the proprioceptive feedback gain in the voltage dynamics. $i_p$ makes the output voltage grow (decrease) as the crawler contracts (extends). Thus, the resulting closed-loop electromechanical interaction is of negative feedback kind. Accordingly, the control input $f$ is proportional to the voltage produced by
\begin{equation}
      c \, \dot V  =  -\kappa \,  V^3 + \rho \, V - k_p \, s + i, 
      \label{eq:voltage_dynamics}
\end{equation}
where $c$ is a capacitance and $i$ is an exogenous input current. The circuit representation of the dynamics \eqref{eq:voltage_dynamics} is in 
Fig. \ref{fig:crawler_schematics}\textbf{B}.

\begin{rem}
\label{rem:current_i}
The external current $i$ in \eqref{eq:voltage_dynamics} could be used, for example, to initiate crawling from equilibrium or to incorporate extraproprioceptive signals and contextual information into the controller dynamics. However, as we show in this work, it is not necessary to generate peristaltic waves in the closed-loop system and thus, for simplicity, we set $i \equiv 0$. 
\end{rem}

\subsection{The excitable crawler}
\label{subsec:excitable_crawler}

 We non-dimensionalize\footnote{ The benefits of nondimensionalization include a reduction in the dimension of the parameter space, identifying dominant terms in the dynamics,
and the definition of \textit{dimensionless groups} with physical interpretation. 
By Buckingham's $\Pi$-theorem \cite[Ch. 1]{IsaakovichBarenblatt2006}, the dimensionless counterpart of the electromechanical closed-loop dynamics has \textit{8 dimensionless groups}. } the closed-loop electromechanical dynamics.  We define \textit{dimensionless variables}
\begin{equation}
    \nd{t} := t / t_* , \; \nd{s} := s/l_*, \text{ and } \nd{V} := V / V_*,
\label{eq:dimensionless_variables}
\end{equation} 
 where $V_*$ denotes the characteristic scale of the voltage.
Accordingly, the  dimensionless dynamics of the electromechanical closed-loop are
\begin{subequations}
\begin{small}
    \begin{align}
        \nd{V}^{\prime}  &= - \pi_{\nd{c}} \,  \nd{V}^3 +\pi_{\nd{l}} \, \nd{V} - \pi_{\nd{s}} \, \nd{s}, \label{eq:V_dynamics}\\
  \nd{v}_{\text{com}}^{\prime} &=  - \frac{\pi_{\nd{f}}}{2} \, \Big( \sigma_{\pi}\Big(\nd{v}_{\text{com}} - \frac{\nd{v_s}}{2}\Big) + \sigma_{\pi}\Big(\nd{v}_{\text{com}} + \frac{\nd{v_s}}{2}\Big)  \Big), \label{eq:CoM_dynamics}\\
  \nd{s}^{\prime} & = \nd{v_s}, \label{eq:s_dynamics}\\
   \nd{v_s}^{\prime} &=  \pi_{\nd{f}} \Big( \sigma_{\pi} \Big(\nd{v}_{\text{com}} - \frac{\nd{v_s}}{2}\Big) - \sigma_{\pi} \Big(\nd{v}_{\text{com}} + \frac{\nd{v_s}}{2}\Big) \Big)   - \nd{s}  - 2  \zeta  \nd{v_s}   + 2 \pi_{\nd{v}} \nd{V}, \label{eq:vs_dynamics}
\end{align}
\end{small}
where $\nd{v_{\text{com}}}$  denotes the speed of the crawler's center of mass and $\nd{v_s}$ the strain rate of the crawler's body. $(\cdot)^{\prime}$ denotes $\mathrm{d}(\cdot)/\mathrm{d}\nd{t}$ and the function $\sigma_{\pi}(\cdot)$ is 
\begin{small}
\begin{equation}
    \sigma_{\pi}(\nd{u}') := \frac{\tanh(\pi_{\epsilon} \nd{u}' + \nd{n_f}) - \tanh(\nd{n_f})}{1 + \tanh{(\nd{n_f})}}.
\label{eq:dimensionless_friction}
\end{equation}
\end{small}
\label{eq:dimensionless_electromechanical_dynamics_movingFrame}
\end{subequations}
The definitions of the dimensionless groups in \eqref{eq:dimensionless_electromechanical_dynamics_movingFrame} are provided in Table \ref{table:system_params}. We define the state as $\boldsymbol{\nd{x}}(\nd{t}) := [\nd{V}\nd{(t)} \; \nd{v}_{\text{com}}\nd{(t)} \; \nd{s}\nd{(t)} \; \nd{v_s}\nd{(t)}]^{\top}$ and succinctly write the dynamics \eqref{eq:dimensionless_electromechanical_dynamics_movingFrame} as $\boldsymbol{\nd{x}}^{\prime} = \boldsymbol{f}(\boldsymbol{\nd{x}})$. We refer to the closed-loop system \eqref{eq:dimensionless_electromechanical_dynamics_movingFrame} as \textit{the excitable crawler}\footnote{By \textit{excitability} we refer to the characteristic behavior of mixed-feedback systems, in which negative and positive feedback coexist but are separated in range, timescale, or spatial scale \cite{Sepulchre2018}.}. The following are two useful facts related to the excitable crawler dynamics:
\begin{itemize}
    \item \textbf{Equilibria.} The fixed points of \eqref{eq:dimensionless_electromechanical_dynamics_movingFrame} can be characterized explicitly. They consist of the central fixed point $\boldsymbol{\mathsf{x}}_0 = \mathbf{0}$ and two symmetric fixed points $ \boldsymbol{\nd{x}}_{\pm}$: 
    \begin{small}
    \begin{equation}
             \boldsymbol{\nd{x}}_{\pm} =
             \sqrt{\frac{\pi_{\nd{l}} - 2 \pi_{\nd{V}} \pi_{\nd{s}}}{\pi_{\nd{c}}}}
        \begin{bmatrix}
             \pm 1 & 
            0 & 
           \pm  2 \pi_{\nd{V}}  & 
            0
        \end{bmatrix}^\top.
        \label{eq:FPs_sym}
         \end{equation}
         \label{eq:FPs}
    \end{small}
     \item \textbf{Equivariance.} The dynamics \eqref{eq:dimensionless_electromechanical_dynamics_movingFrame} are $g$-symmetric, with  $g:= \mathrm{diag} \big( [-1,\, 1,\, -1,\, -1]\big) \in \mathbf{O}(4)$. 
\end{itemize}
 By definition, the symmetry in the excitable crawler dynamics implies that if $\boldsymbol{\nd{x}}$ is a trajectory, so is $g \boldsymbol{\nd{x}}$. This manifests, for example, in the symmetric fixed points satisfying $\boldsymbol{\nd{x}}_+ = g \boldsymbol{\nd{x}}_-$, in the bifurcations happening at these fixed points (see \S \ref{sec:organizing_bifurcations}), and in the global oscillations that lead to a crawling gait in the excitable crawler (see \S\ref{sec:GSP}).

\begin{table}[h]
\begin{center}
\caption{ \begin{small} (Left) Dimensional parameters of the electromechanical closed-loop crawler and their base units: mass ($M$), length ($L$), time ($T$), and current ($I$). (Right) Definition of the dimensionless groups in \eqref{eq:dimensionless_electromechanical_dynamics_movingFrame}. $\zeta$ is the damping ratio of the crawler; $\pi_{\nd{f}}$ and $\pi_{\nd{V}}$ are the ratios of the frictional and feedback actuator forces to the elastic force, respectively. $\pi_{\epsilon}$ and $\nd{n_f}$ are the slope and anisotropy parameters of the dimensionless friction model.
 \end{small}}
 \label{table:system_params}
 \begin{small}
\begin{tabular}{|c | c|} 
 \hline
 Dimensional & Base    \\
 parameter & Units  \\ [0.5ex] 
 \hline\hline
 $m$ & $M$  \\ 
 \hline
 $\ell$ & $L$  \\
 \hline
 $k$ &  $M T^{-2}$   \\
 \hline
 $b$ &  $M T^{-1}$ \\
 \hline
 $A_\sigma$ & $M L T^{-2}$  \\
  \hline
$k_v$ &  $T I L^{-1}$  \\ 
 \hline
$c$ &  $M^{-1} L^{-2} T^4 I^2$ \\ 
 \hline
$\kappa$ & $M^{-3} L^{-6} I^4 T^9$ \\ 
 \hline
 $\rho$ & $I^2 T^3 M^{-1} L^{-2}$ \\ 
  \hline
$k_p$ & $I L^{-1}$ \\ 
 \hline
 $\epsilon_f$ & $L T^{-1}$ \\
 \hline
\end{tabular}
\quad
\begin{tabular}{|c|}
\hline 
Dimensionless\\
Group \\
\hline \hline
 $ \zeta  := b/\sqrt{2mk}$ \\
 \hline
 $\pi_{\nd{f}}  := A_\sigma/( 2\, k \, l_*)$ \\
 \hline
 $\pi_{\nd{v}}  := \frac{1}{2} \,  k_v V_* / (k \, l_*)$ \\
 \hline
 $\pi_{\epsilon} := l_*/(t_* \epsilon_f) $ \\
 \hline
     $\nd{n_f}$ \\
    \hline
 $\pi_{\nd{c}} := \kappa \, V_*^2 t_* / c  $ \\
 \hline
 $\pi_{\nd{l}} := \rho t_*/ c $ \\
 \hline
  $\pi_{\nd{s}}  := k_p \, l_*  t_* / (c \, V_* )$ \\
  \hline
\end{tabular}
 \end{small}
\end{center}
\end{table}

\vspace{-.75cm}
\section{Organizing bifurcations of resting and crawling behaviors}
\label{sec:organizing_bifurcations}

\begin{figure}[H]
    \centering
    \includegraphics[width=\linewidth]{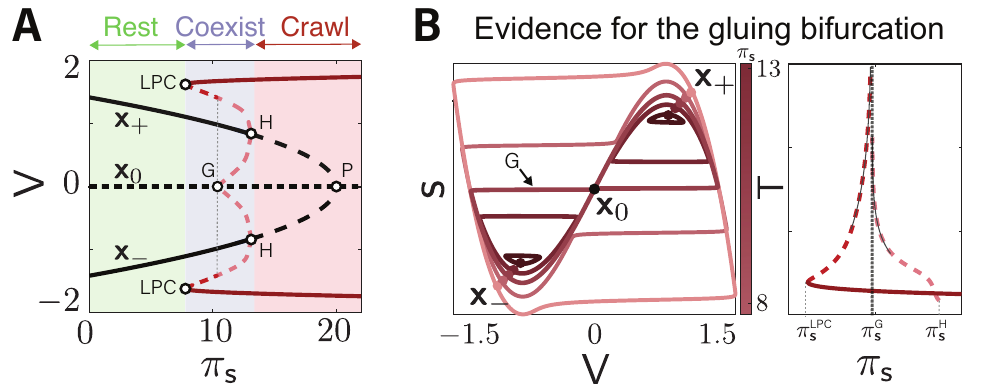}
\caption{
\begin{small}
Bifurcation structure and evidence for a symmetry-organized gluing bifurcation in the excitable crawler. Parameter values are $
\pi_\nd{c}=10,
\pi_\nd{l}=20,
\pi_\nd{V}=0.5,
\pi_\nd{f}=2.5,
\zeta=0.5,
\pi_\epsilon=10 \text{ and }
\nd{n}_\nd{f}=1.5$.
\textbf{(A)} Bifurcation diagram with $\pi_\nd{s}$ as continuation parameter.
Black curves denote equilibria; red curves show the minimum and maximum values of $\nd{V}$ along periodic orbits.
Solid (dashed) curves denote stable (unstable) solutions.
The equilibria $\boldsymbol{\nd{x}_{\pm}}$ lose stability through subcritical Hopf bifurcations $\mathrm{H}$, from which two unstable local periodic orbits emerge.
As $\pi_\nd{s}$ decreases, these cycles approach the saddle $\boldsymbol{\nd{x}}_0$ and undergo a global gluing bifurcation at $\mathrm{G}$, continuing as a single large unstable periodic orbit.
This branch collides with the stable large-amplitude cycle at the limit point of cycles $\mathrm{LPC}$.
The nonzero equilibrium branches meet the central branch at the pitchfork bifurcation $\mathrm{P}$.
Background shading identifies different regimes: resting (green), coexistence (blue), and crawling (red); in the coexistence regime, stable resting and crawling solutions coexist.
\textbf{(B)}
Evidence for the gluing bifurcation.
(Left)
Representative periodic orbits in the $(\mathsf{V},\mathsf{s})$-plane, colored by $\pi_\nd{s}$ according to the colorbar.
The $\mathbb{Z}_2$ symmetry  organizes the simultaneous homoclinic connections to $\boldsymbol{\sf{x}}_0$ at $\mathrm{G}$.
(Right)
Period $\mathsf{T}$ of the periodic-orbit branches as a function of $\pi_\nd{s}$.
The two-sided growth of the period as $\pi_\nd{s} \to\pi_\nd{s}^{\mathrm{G}}$ is consistent with the logarithmic divergence expected near a homoclinic gluing bifurcation (gray dotted curves are logarithmic fits).
Solid red branch denotes the stable large-amplitude periodic crawling orbit.
\end{small}
}
    \label{fig:bifurcation_diagram}
\end{figure}

The ability of the excitable crawler to sustain peristaltic waves and
generate locomotion depends on the bidirectional sensorimotor coupling
between the controller and the crawler's body. For example, if either
$\pi_\nd{s}$ or $\pi_\nd{V}$ vanishes in \eqref{eq:dimensionless_electromechanical_dynamics_movingFrame}, this coupling is broken and the
crawler cannot generate endogenous locomotion.
We call asymptotically stable equilibria
\textit{resting solutions}, since their
center-of-mass velocity vanishes. We call orbitally stable periodic orbits with nonzero average center-of-mass velocity \textit{crawling solutions}. 
We first characterize analytically the local bifurcations of the
equilibria as the sensorimotor gain $\pi_\nd{s}$ varies. We then complement
these results through numerical continuation of the periodic orbit
branches, revealing the global bifurcation structure connecting
resting and crawling solutions---summarized in Fig. \ref{fig:bifurcation_diagram}\textbf{A}. We introduce the following assumption
\begin{assumption}[Parameter ranges]
\label{ass:parameter_ranges}
    $\pi_\nd{s}>0, \; \pi_\nd{V}>0, \; \pi_\epsilon>0, \; \pi_\nd{c}>0, \; \pi_\nd{f}>0, \; \pi_\nd{l}>0, \; \nd{n_f} > 0, \; \zeta > 0$.
\end{assumption}

For convenience, we define $\gamma:= \pi_\nd{f} \sigma'_\pi(0) + 2 \zeta$ and a critical value of the sensorimotor gain
   \begin{small}
        \begin{equation}
                \pi_{\nd{s}}^\nd{H}   = \frac{1}{\pi_{\nd{V}}} \bigg(
                \frac{1}{12} \gamma + \frac{1}{36 \gamma} + \frac{1}{3} \pi_{\nd{l}} - \frac{1}{2}\sqrt{\frac{1}{18^2 \gamma^2 } + \frac{1}{36 }\gamma^2 + \frac{2 \pi_\nd{l}}{27 \gamma} -\frac{5}{54}} \, \bigg). 
                \label{eq:parameter_hopf_1}
        \end{equation}
    \end{small}
We show next that for low values of $\pi_\nd{s}$ the symmetric fixed points are locally asymptotically stable.
\begin{prop}[Stability of the symmetric resting equilibria]
\label{prop:bistability}
Let Assumption \ref{ass:parameter_ranges} hold and further assume that
$1<\gamma<\pi_\nd{l}$. Then, the symmetric fixed points $\boldsymbol{\mathsf{x}}_{\pm}$ as given in \eqref{eq:FPs_sym} are locally asymptotically stable for $0 < \pi_\nd{s} < \pi_\nd{s}^\nd{H}$, where $\pi_\nd{s}^\nd{H}$ is as defined in \eqref{eq:parameter_hopf_1}. 
\end{prop}
\begin{proof}
    See \hyperref[subsec:proof_bistability]{Appendix I.B}. 
\end{proof}

The stability of $\boldsymbol{\nd{x}}_\pm$ is lost in a paired Hopf bifurcation.
\begin{thm}[Paired Hopf bifurcation at $\boldsymbol{\nd{x}}_{\pm}$]
    \label{thm:equivariant_Hopf}
 Consider the closed-loop excitable crawler system \eqref{eq:dimensionless_electromechanical_dynamics_movingFrame} and the bifurcation parameter $\pi_{\nd{s}}$. Let Assumption \ref{ass:parameter_ranges} hold and further assume
    that  $1< \gamma < \pi_\nd{l}. $ 
Then, the fixed points $\boldsymbol{\nd{x}_{\pm}}$ 
   simultaneously undergo a  Hopf bifurcation at $\pi_\nd{s} = \pi_\nd{s}^\nd{H}$, with $\pi_\nd{s}^\nd{H}$ as defined in \eqref{eq:parameter_hopf_1}. Furthermore, whenever the first Lyapunov coefficient $l_1$ as given in \eqref{eq:l1} satisfies $l_1>0$ (resp., $l_1<0$), the Hopf bifurcation is subcritical (resp., supercritical).
\end{thm}
\begin{proof}
See \hyperref[subsubsec:proof_equivariant_Hopf]{Appendix I.C}.
\end{proof}

The paired Hopf bifurcations mark the loss of local stability of the
symmetric resting states as $\pi_\nd{s}$ increases through $\pi_\nd{s}^\nd{H}$.
For the parameter values considered in Fig. \ref{fig:bifurcation_diagram},
the first Lyapunov coefficient is positive; hence, the bifurcations are
subcritical and two symmetry-related unstable small-amplitude
periodic orbit branches emerge for $\pi_\nd{s}<\pi_\nd{s}^\nd{H}$. Local orbital
stability of the periodic orbit branches is determined by the nontrivial Floquet multipliers
\cite[\S1.5.2]{Kuznetsov2023}. Stronger
basin-level guarantees could potentially be sought through transverse
contraction theory \cite{Manchester2013}. As $\pi_\nd{s}$ increases
beyond $\pi_\nd{s}^\nd{H}$, the nonzero equilibrium branches persist and
subsequently meet the central equilibrium $\boldsymbol{\nd{x}}_0$ in the pitchfork
bifurcation characterized next.

\begin{thm}[Pitchfork bifurcation at $\boldsymbol{\nd{x}_0}$]
\label{thm:pitchfork_crawler}
    Consider the closed-loop excitable crawler system \eqref{eq:dimensionless_electromechanical_dynamics_movingFrame} and the bifurcation parameter $\pi_{\nd{s}}$. Let Assumption \ref{ass:parameter_ranges} hold and further assume that $\gamma \pi_\nd{l} > 1$. 
    Then, the excitable crawler \eqref{eq:dimensionless_electromechanical_dynamics_movingFrame} undergoes a pitchfork bifurcation at the central equilibrium $\boldsymbol{\nd{x}}_0$ at:
    \begin{small}
    \begin{equation}
    \pi_\nd{s}^\nd{P} = \frac{\pi_\nd{l}}{2 \, \pi_\nd{V}}.
        \label{eq:pitchfork_bifurcation_value}
    \end{equation}
    \end{small}
    The non-trivial equilibrium branches exist for $\pi_\nd{s}<\pi_\nd{s}^\nd{P}$. Locally near the bifurcation, these branches are unstable on the one-dimensional center manifold.
\end{thm}
\begin{proof}
    See \hyperref[subsubsec:pitchfork_proof]{Appendix I.D}.
\end{proof}
The analytical results regarding local bifurcations of the equilibria are complemented by numerically continuing the
equilibrium and periodic orbit branches using
\textsc{MatCont} \cite{Dhooge2003}; see
Fig. \ref{fig:bifurcation_diagram}. The unstable local cycles born at the
paired subcritical Hopf bifurcations grow as $\pi_\nd{s}$ decreases and
approach the central saddle $\boldsymbol{\nd{x}}_0$. At
$\pi_\nd{s}=\pi_\nd{s}^\nd{G}$, the numerical continuation
provides evidence for simultaneous homoclinic connections to $\boldsymbol{\nd{x}}_0$,
organized by reflection symmetry in a \textit{gluing bifurcation}. They subsequently continue as a single large unstable
periodic orbit, which collides with the stable large-amplitude crawling
orbit at the \textit{limit point of cycles}
$\pi_\nd{s}=\pi_\nd{s}^{\nd{LPC}}$.
The remainder of the paper focuses on the stable
crawling orbit and on the slow--fast geometry underlying its
peristaltic gait.

\section{Separation of Timescales
}
\label{sec:GSP}

We turn our attention back to the characteristic voltage scale $V_*$. Dimensional estimates identify a representative regime with separated
electrical and mechanical timescales, which we leverage to analyze the
excitable crawler using Geometric Singular Perturbation (GSP) theory \cite{Jones1995}.

\subsection{Fast \& slow dynamics}

A \textit{slow-fast system} is a system of differential equations in which some variables
have their derivatives with larger magnitude than others, leading to a system with multiple time scales. In the dimensional model of \S\ref{subsec:excitable_crawler}, we expect $V_* \sim 10^{-1}$ volts, $\ell \sim 10^{-1}$meters, $c \sim 10^{-6}$ farads, and $t_* \sim 10^{-1}$s.  
We set $\kappa \sim 10$ and $\rho, k_p \sim 10^{-1}$ (all in SI units) to yield reasonable strain levels in the crawler. Then, $\pi_{\nd{l}}, \pi_{\nd{c}}, \pi_{\nd{s}} \sim 10^4$. Consequently,  we define $V_* := 10^2 \sqrt{c/(t_* \, \kappa)}$. Since entries in the dimensionless state have been normalized to be of order 1 and dimensionless groups in the crawler dynamics are also expected to be of order 1, the relatively large values of $\pi_{\nd{l}}, \pi_{\nd{c}}, \pi_{\nd{s}}$ imply that for this representative parameter regime the voltage dynamics \eqref{eq:V_dynamics} are orders of magnitude faster than the crawler's body dynamics: there is a separation of timescales between electrical (controller) and crawler's body (mechanical) dynamics. Next, the timescale separation is made explicit in the excitable crawler dynamics \eqref{eq:dimensionless_electromechanical_dynamics_movingFrame} yielding a 
slow-fast-system (1 fast, 3 slow).
For notational convenience, define 
$\sigma_\pi^{\pm}:= \sigma_\pi(\nd{v}_{\nd{com}} \pm \nd{v_s}/2)$. Then,

\underline{\textit{Slow dynamics.}} $\nd{v}_{\text{com}}(\mathsf{t}), \nd{s}(\mathsf{t}), \nd{v_s}(\mathsf{t})$ are the \textit{slow state variables} and $\nd{t}$ is the \textit{slow timescale}.
\begin{small}
\begin{subequations}
    \begin{align}
        \varepsilon \, \nd{V}^{\prime}  &= - \pi_{\nd{c}}^{(\varepsilon)} \,  \nd{V}^3 +\pi_{\nd{l}}^{(\varepsilon )} \, \nd{V} - \pi_{\nd{s}}^{(\varepsilon )} \, \nd{s}, \\
  \nd{v}_{\text{com}}^{\prime} &=  - \frac{\pi_{\nd{f}}}{2} \, \big( \sigma_{\pi}^- + \sigma_{\pi}^+\big), \\
  \nd{s}^{\prime} & = \nd{v_s}, \\
   \nd{v_s}^{\prime} &=  \pi_{\nd{f}} \big( \sigma_{\pi} ^{-} - \sigma_{\pi}^+ \big)   - \nd{s}  - 2 \, \zeta \, \nd{v_s}   + 2 \pi_{\nd{V}} \nd{V},
\end{align}
\label{eq:slowDynamics}
\end{subequations}
\end{small}
where $\varepsilon = 10^{-4}$ and $\pi_{\nd{i}}^{(\varepsilon)} := \varepsilon \, \pi_{\nd{i}}$ (with $\nd{i}= \nd{l}, \nd{c}, \nd{s})$---which are dimensionless groups of order 1.

 \textit{\underline{Fast dynamics}.} We define the \textit{fast timescale} $\nd{T}$ as
$
   \nd{T} := \nd{t}/\varepsilon. 
$
$\nd{V}(\mathsf{T})$ is the \textit{fast state variable}. 
A change in timescale yields
\begin{small}
\begin{subequations}
    \begin{align}
        \dot{\nd{V}} &= - \pi_{\nd{c}}^{(\varepsilon)} \,  \nd{V}^3 +\pi_{\nd{l}}^{(\varepsilon )} \, \nd{V} - \pi_{\nd{s}}^{(\varepsilon )} \, \nd{s}, \\
  \dot{\nd{v}}_{\text{com}} &=  - \varepsilon \frac{\pi_{\nd{f}}}{2} \, \big( \sigma_{\pi} ^-  + \sigma_{\pi}^+ \big), \\
  \dot{\nd{s}} & = \varepsilon \, \nd{v_s}, \\
   \dot{\nd{v}}_{\nd{s}} &=  \varepsilon \big(\pi_{\nd{f}} ( \sigma_{\pi} ^- - \sigma_{\pi}^+ )  - \nd{s}  - 2 \, \zeta \,  \nd{v_s}   + 2 \pi_{\nd{V}} \nd{V}\big), 
\end{align}
\label{eq:fastDynamics}
\end{subequations}
\end{small}
where by a slight abuse of notation we denoted $\mathrm{d} \nd{x}/\mathrm{d} \nd{T}$ by $ \dot{\nd{x}}$. 

The slow \eqref{eq:slowDynamics} and fast \eqref{eq:fastDynamics} dynamics are amenable to Geometric Singular Perturbation analysis \cite{Jones1995}. 
Considering the singular limit $\varepsilon \to 0$, the independence of the two distinct timescales---slow and fast---in the system makes the dynamics simpler to analyze. Fenichel theory \cite{Fenichel1979} ensures the validity of the analysis away from the singular limit for normally hyperbolic critical manifolds and sufficiently small $\varepsilon$.

\textit{\underline{Singularly perturbed slow dynamics}.} By setting $\varepsilon = 0$ in 
\eqref{eq:slowDynamics} we obtain the \textit{reduced system}, also called the \textit{slow subsystem}. This system is a differential algebraic system describing the evolution of the slow variables $\nd{v}_{\text{com}}, \nd{s}$, and $\nd{v_s}$:  
\begin{small}
\begin{subequations}
    \begin{align}
        0  &= - \pi_{\nd{c}}^{(\varepsilon)} \,  \nd{V}^3 +\pi_{\nd{l}}^{(\varepsilon )} \, \nd{V} - \pi_{\nd{s}}^{(\varepsilon )} \, \nd{s}, \label{eq:slow_manifold}\\
  \nd{v}_{\text{com}}^{\prime} &=  - \frac{\pi_{\nd{f}}}{2} \, ( \sigma_{\pi} ^- + \sigma_{\pi}^+ ), \\
  \nd{s}^{\prime} & = \nd{v_s}, \\
   \nd{v_s}^{\prime} &=  \pi_{\nd{f}} ( \sigma_{\pi} ^- - \sigma_{\pi}^+ )  - \nd{s}  - 2 \zeta \nd{v_s}   + 2 \pi_{\nd{V}} \nd{V}.
\end{align}
\label{eq:slowDynamics_singularPerturbation}
\end{subequations}
\end{small}
The set
\begin{equation}
    \mathcal{S}_0:= \big\{(\nd{V}, \nd{v}_{\nd{com}}, \nd{s}, \nd{v}_\nd{s}) \; |  \;   \pi_{\nd{c}}^{(\varepsilon)} \,  \nd{V}^3 -\pi_{\nd{l}}^{(\varepsilon )} \, \nd{V} + \pi_{\nd{s}}^{(\varepsilon )} \, \nd{s} = 0 \big\},
\label{eq:critical_manifold}
\end{equation}
consisting of the equilibria of the fast dynamics, is known as the \textit{critical manifold}\footnote{A subset $\mathcal{S}_h \subseteq \mathcal{S}_0$ is called \textit{normally hyperbolic} if all $(\nd{V}, \nd{v}_\nd{com}, \nd{s}, \nd{v}_\nd{s}) \in \mathcal{S}_h$ are hyperbolic equilibria of the layer problem, that is, $\partial_\nd{V} \big( -\pi_\nd{c} \nd{V}^3 + \pi_\nd{l} \nd{V} - \pi_\nd{s} \nd{s} \big) \neq 0$. We call a normally hyperbolic subset $\mathcal{S}_a \subseteq \mathcal{S}_0$ (resp., $\mathcal{S}_r \subseteq \mathcal{S}_0$) \textit{attracting} (resp., \textit{repelling}) if $\partial_\nd{V} \big( -\pi_\nd{c} \nd{V}^3 + \pi_\nd{l} \nd{V} - \pi_\nd{s} \nd{s} \big) <0$ for all $ (\nd{V}, \nd{v}_\nd{com}, \nd{s}, \nd{v}_\nd{s}) \in \mathcal{S}_a$ (resp. $>0$, for all $ (\nd{V}, \nd{v}_\nd{com}, \nd{s}, \nd{v}_\nd{s}) \in \mathcal{S}_r$). 
We use analogous terminology when referring to subsets of the \textit{slow manifold} $\mathcal{S}_\varepsilon$ instead. We also note that $\mathcal{S}_\varepsilon$ is, in general, not unique---but all representations lie $\varepsilon$-exponentially close from each other.}. 

\textit{\underline{Singularly perturbed fast dynamics}.}
$\nd{V}$ evolves rapidly while $\nd{v_{\text{com}}}, \nd{s}$, and $\nd{v_s}$ remain constant. This behavior is described by the \textit{layer problem} or \textit{fast subsystem}:
\begin{small}
\begin{subequations}
    \begin{align}
        \dot{\nd{V}} &= - \pi_{\nd{c}}^{(\varepsilon)} \,  \nd{V}^3 +\pi_{\nd{l}}^{(\varepsilon )} \, \nd{V} - \pi_{\nd{s}}^{(\varepsilon )} \, \nd{s},\\
  \dot{\nd{v}}_{\text{com}} &=  0, \;\; \dot{\nd{s}}  = 0, \;\;\;  \nd{\dot{v}_s} = 0.
\end{align}
\label{eq:fastDynamics_singularlyPerturbed}
\end{subequations}
\end{small}

\vspace{-1cm}
\subsection{The geometry of the critical manifold}

We analyze how the slow--fast structure of the excitable crawler shapes the periodic orbit branches identified in
Fig. \ref{fig:bifurcation_diagram}. In the singularly perturbed regime, the
paired Hopf bifurcations occur close to the folds of the
$\mathsf{S}$-shaped critical manifold. The unstable cycles born at these bifurcations can therefore be influenced by the geometry of the attracting and
repelling branches of the manifold and by the folded singularities located at
its folds. The main geometric features are summarized in the following proposition.

\begin{prop}[Critical  manifold geometry and Hopf bifurcations near the folds]
\label{prop:canards} Consider the excitable crawler dynamics \eqref{eq:dimensionless_electromechanical_dynamics_movingFrame} in the singularly perturbed regime, with critical manifold $\mathcal{S}_0$ as defined in \eqref{eq:critical_manifold}. Then, \\
    \indent a) The critical manifold has folds 
    \begin{equation}
    \mathcal{F}:= \big\{ (\nd{V}, \nd{v}_\nd{com}, \nd{s}, \nd{v}_\nd{s})  \, | \, \nd{V} = \nd{V}_{\pm}^{\nd{F}}, \, \nd{s} = \nd{s}_{\pm}^{\nd{F}} \big\},
    \label{eq:folds}
\end{equation}
where $\nd{V}_{\pm}^{\nd{F}} = \pm \sqrt{\frac{\pi_\nd{l}^{(\varepsilon)}}{3 \, \pi_{\nd{c}}^{(\varepsilon)}}}$ and $\nd{s}_{\pm}^{\nd{F}} = \pm \frac{2 \big(\pi_{\mathsf{l}}^{(\varepsilon)}\big)^{3/2} }{3 \sqrt{3} \big(\pi_{\mathsf{c}}^{(\varepsilon)}\big)^{1/2} \pi_{\mathsf{s}}^{(\varepsilon)}}$. \\
\indent b) The reduced (slow) dynamics admit  folded singularities located at
   \begin{equation}
   M_\mathcal{F} := \big \{(\nd{V}, \nd{v}_\nd{com}, \nd{s}, \nd{v}_\nd{s}) \; | \; \nd{V} = \nd{V}_\pm^{\nd{F}}, \, \nd{s} = \nd{s}_\pm^{\nd{F}},  \, \nd{v}_\nd{s} = 0 \big\}.
   \label{eq:folded_singularities}
   \end{equation}
   \indent c) If 
    $
        0 < \pi_\nd{s}< \frac{\pi_\nd{l}}{3 \pi_\nd{V}} 
    $
    the folded singularities in $M_{\mathcal F}$ are of saddle type. \\
    \indent d) 
Let $\boldsymbol{\mathsf{x}}_{\pm}(\pi_{\mathsf{s}})$ denote the symmetric equilibria \eqref{eq:FPs_sym} undergoing a Hopf bifurcation at
$\pi_{\mathsf{s}}=\pi_{\mathsf{s}}^{\mathsf{H}}$.
As $\varepsilon \to 0^+$, these equilibria satisfy
$
\mathsf{V}_{\pm}(\pi_{\mathsf{s}}^{\mathsf{H}})
=
\mathsf{V}_{\pm}^{\mathsf{F}}
+
\mathcal{O}(\varepsilon^{1/2}) \text{ and }
\mathsf{s}_{\pm}(\pi_{\mathsf{s}}^{\mathsf{H}})
=
\mathsf{s}_{\pm}^{\mathsf{F}}
+
\mathcal{O}(\varepsilon^{1/2}).
$
Moreover, in the slow time scale $\mathsf{t}$, the Jacobian of the full system evaluated at
$\boldsymbol{\mathsf{x}}_{\pm}(\pi_{\mathsf{s}}^{\mathsf{H}})$
has a pair of purely imaginary eigenvalues
$
\lambda_{2,3}(\varepsilon)
=
\pm \mathbf{i} \,\omega(\varepsilon) \text{ with }
\omega(\varepsilon) 
=
\Theta(\varepsilon^{-1/4})
$.
\end{prop}
\begin{proof}
See \hyperref[sec:Appendix_II]{Appendix II}.
\end{proof}

Proposition \hyperref[prop:canards]{V.1d)} shows that, in the singular  regime,
the paired Hopf bifurcations occur asymptotically close to the folds
of the critical manifold, where normal hyperbolicity is lost. Thus,
the small-amplitude periodic orbits born at Hopf emerge precisely in
the region where the attracting and repelling slow-manifold sheets
meet and can be influenced by the surrounding slow-fast geometry. Moreover, Proposition \hyperref[prop:canards]{V.1c)}  shows that the corresponding folded
singularities are of saddle type, which can organize passage through a
fold from an attracting sheet of the slow manifold to a repelling
sheet. The associated singular canards\footnote{ A \textit{canard} is a solution of a singularly perturbed dynamical system that follows an attracting slow-manifold sheet through a fold region and then remains close to a repelling sheet for a finite time \cite{Wechselberger2013}.  Given a singularly perturbed system with a folded critical manifold $\mathcal{S}_0 = \mathcal{S}_a \cup \mathcal{F} \cup \mathcal{S}_{r}$, a trajectory of the reduced dynamics that has the ability to cross in finite time from the $\mathcal{S}_a$ branch to the $\mathcal{S}_{r}$ branch via a folded singularity is called a \textit{singular canard}. 
For sufficiently small $\varepsilon>0$, a \textit{maximal canard} is a trajectory contained in the intersection of the
corresponding attracting and repelling branches of the slow manifold.
} can
therefore persist, for sufficiently small $\varepsilon>0$, as maximal
canards connecting attracting and repelling slow-manifold sheets \cite{Wechselberger2013}.
This provides a geometric mechanism by which the unstable periodic
orbits born at the paired Hopf bifurcations can acquire canard-like geometry: after entering the fold region, they may remain close to a
repelling slow-manifold sheet for a finite time before departing.
The numerically continued unstable cycles in Fig. \ref{fig:bifurcation_diagram}\textbf{B} exhibit shapes
consistent with this picture. Reflection symmetry produces two such
branches, associated with the symmetric equilibria $\boldsymbol{\nd{x}}_{\pm}$, whose
global organization is revealed by the continuation in Fig. \ref{fig:bifurcation_diagram}.

\begin{figure}[h]
    \centering
    \includegraphics[width=\linewidth]{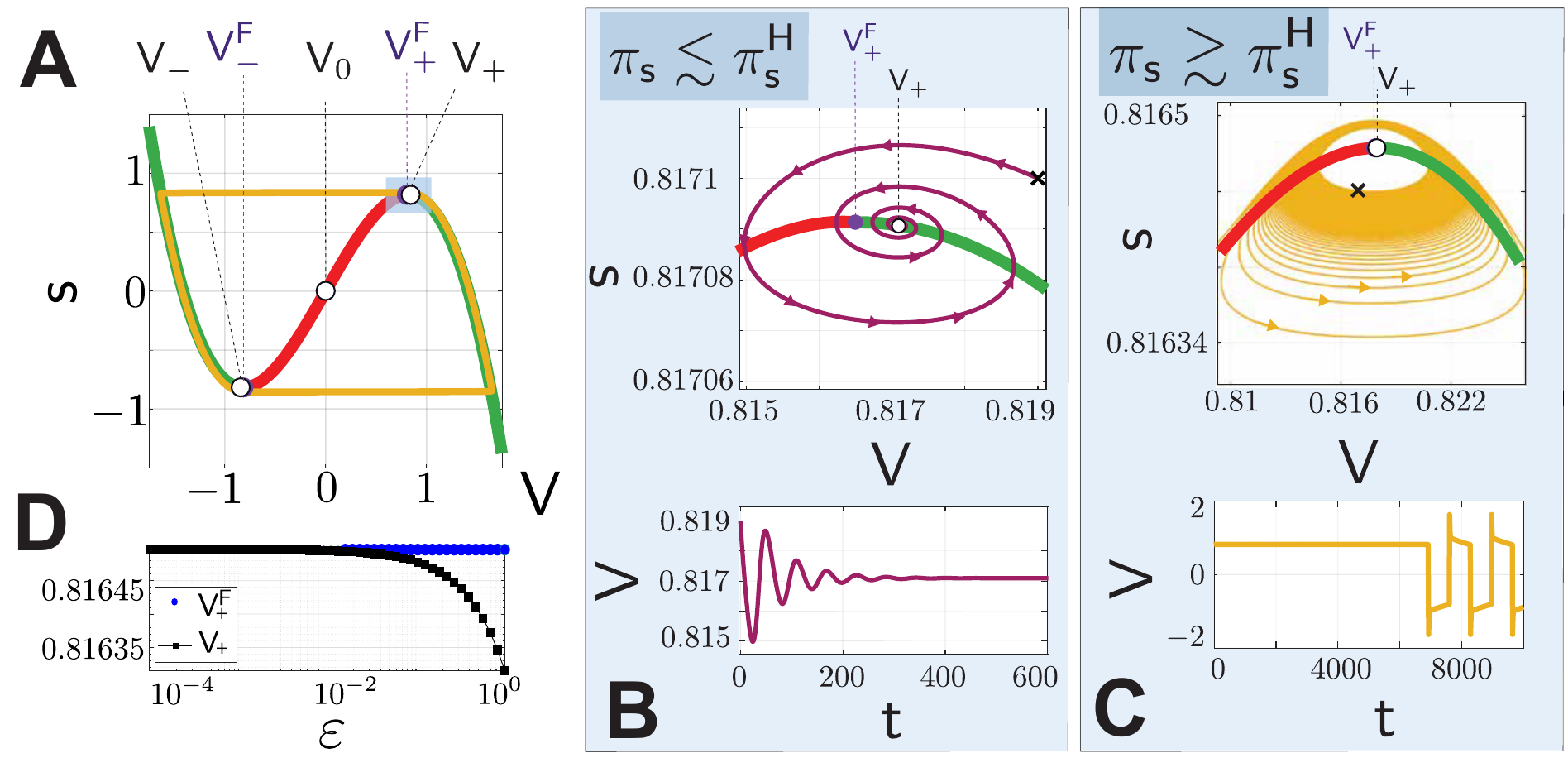}
    \caption{
    \begin{small} Behavior of the excitable crawler in a neighborhood of the symmetric fixed point $\boldsymbol{\nd{x}}_+$  projected into the $\nd{V}$-$\nd{s}$ plane for two different values of the sensorimotor gain $\pi_\nd{s}$, immediately below and above the Hopf bifurcation at $\pi_\nd{s}^\nd{H}$ in the singularly perturbed regime. $\boldsymbol{\nd{x}}_+$ is located close to the fold of the critical manifold.
    \textbf{(A)} Critical manifold $\mathcal{S}_0$, with the attracting (repelling) sheets in green (red). White (purple) circles correspond to fixed points (folds). The trajectory in yellow corresponds to global relaxation oscillations. Panels \textbf{(B)} and \textbf{(C)} zoom into the region highlighted in blue in panel \textbf{(A)}, for two different values of $\pi_\nd{s}$. The respective initial condition is represented by a black cross. \textbf{(B)} Convergence to one of the stable symmetric fixed points when $\pi_\nd{s} \lesssim \pi_\nd{s}^\nd{H}$.  \textbf{(C)} The top panel shows the transient trajectory of the excitable crawler at $\pi_\nd{s} \gtrsim \pi_\nd{s}^\nd{H}$ when initialized close to the unstable symmetric fixed point $\boldsymbol{\nd{x}}_+$,
    before the system eventually converges to large-amplitude relaxation oscillations---yellow trajectory. \textbf{(D)} Convergence  $\nd{V}_+ \to \nd{V}_+^\nd{F}$ as $\varepsilon \to 0^+$, as per Proposition \hyperref[prop:canards]{V.1d)}.
    \end{small}
    }
    \label{fig:oscillations_summary}
\end{figure}

\begin{figure}[h]
    \centering
\includegraphics[width=0.48\textwidth]{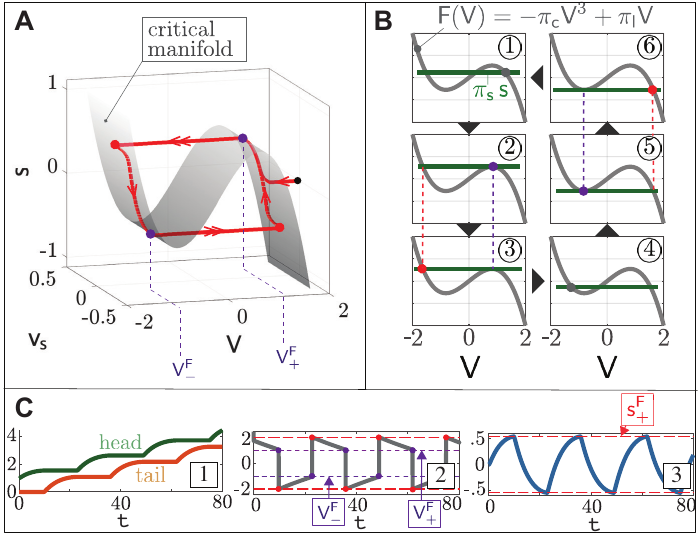}
    \caption{ 
    \begin{small} 
    Relaxation oscillations in the excitable crawler.
    Parameter values: $\zeta = 4.7, \pi_{\nd{f}} =  2.5, \pi_{\nd{V}} = 0.5 , \pi_{\nd{\epsilon}} = 4.7 \cdot 10^3, \nd{n_f} = 1.5, \pi_{\nd{c}} = 10^4, \pi_{\nd{l}} = \pi_{\nd{s}}  = 2 \cdot 10^4$, and initial condition $\nd{V}(0) = 2, \nd{s}(0) = \nd{v_s}(0) = \nd{v_{com}}(0) = 0$. All panels display dimensionless variables.
    \textbf{(A)} Limit cycle in the state variables $\nd{s}, \nd{v_s}, \nd{V}$ in red. Single (double) arrows on the limit cycle correspond to trajectories of the slow (fast) dynamics.  The critical manifold \eqref{eq:critical_manifold} is represented in gray. The initial condition is in black. \textbf{(B)} Schematic of the strain-driven voltage switching. Circled numbers indicate the temporal ordering of the snapshots. The value of the voltage is indicated by the dot in each panel. Purple (red) dots correspond to the value of the voltage right before (after) the switch between attracting branches of the manifold. \textbf{(C)} Time evolution of different state entries of the excitable crawler: \textbf{(C.1)} Displacements of head and tail, as indicated; \textbf{(C.2)} Relaxation oscillations in the voltage. The fast (slow) dynamics correspond to segments going from purple (red) to red (purple) dots; \textbf{(C.3)} Strain in the soft crawler's body. 
    \end{small}} 
    \label{fig:singular_perturbation}
\end{figure}

Numerical experiments 
illustrate the preceding analysis. Fig. \ref{fig:oscillations_summary}
compares trajectories initialized near $\boldsymbol{\nd{x}}_{+}$ immediately
before and after the Hopf bifurcation. For $\pi_\nd{s} \lesssim \pi_\nd{s}^{\nd H}$, the trajectory converges to the
stable resting equilibrium. For
$\pi_\nd{s} \gtrsim \pi_\nd{s}^{\nd H}$, the equilibrium is unstable and the
trajectory leaves its neighborhood before converging to the pre-existing stable large-amplitude crawling orbit---which, in the singularly perturbed regime, takes
the form of a \textit{relaxation oscillation}. As illustrated in
Fig.~\ref{fig:singular_perturbation}, the trajectory alternates between slow
evolution near the attracting sheets of the slow manifold and fast
voltage transitions between them. This mechanism generates periodic
voltage spikes and body deformations, producing the self-sustained peristaltic
crawling gait. Because extension and contraction evolve along
different attracting sheets, the closed-loop system acts as a \textit{hysteretic switch}.

\section{Setting the pace: 
matching neuromechanical scales enables optimal crawls}
\label{sec:setting_the_pace}

Sections \ref{sec:organizing_bifurcations}-\ref{sec:GSP} characterized the emergence of crawling and its slow–fast geometry. We now turn from gait generation to gait performance: how should the internally generated actuation rhythm be tuned relative to the crawler’s mechanics to maximize locomotion speed? It is expected that the corresponding optimality conditions will naturally lead to \textit{synergies between the neural (controller) and body (controlled system) scales of the excitable crawler}. We show, within a relay and harmonic-balance approximation motivated by the singularly perturbed regime, that this optimization leads to a \textit{neuromechanical matching condition} and to optimal crawling at the \textit{body’s natural frequency}.

It has been widely recognized, both in biology and bio-inspired robotics, that \textit{moving at resonance}---i.e., \textit{matching} the  frequency of the periodic control signal to the natural frequency of the body to be controlled---yields multiple advantages, such as minimal metabolic cost, reduction of jitter during movements, and improving cycle-to-cycle repeatability   \cite{Shen2024,Holt1995,Hatsopoulos1996,Goodman2000}.  Experimental observations report that insect-scale soft robots, soft microrobotic transmissions, and other vibration-driven mobile robots often achieve improved speed or efficiency when actuation is tuned near a mechanical resonance 
\cite{Schonebaum2021}. Together with our results, these suggest that matching controller-generated rhythms to body mechanics can be an effective design principle. Indeed, the idea of \textit{entraining} \cite{Jimenez2022} a mechanical system with a neuro-inspired controller has a rich and long history, and  
\cite{Simoni2007} hypothesized that one of the purposes of sensory feedback in rhythmic locomotion is to synergize the nervous and musculoskeletal systems, so that the resulting frequency of movement relates to the body's mechanical resonance.

To make this design principle precise, we maximize the average speed of the crawler’s center of mass using describing-function and harmonic-balance analysis.
This analysis provides guidelines for plant and controller co-design, and for the design of adaptive control schemes. Given the \textit{low-pass} filtering structure of the crawler's body dynamics, the results presented in this section are based on the \textit{describing function analysis}\footnote{An alternative route for our analysis is by following \cite{Chaffey2025}, which introduces the analogue of the describing function method for square waves.}---see \S \ref{sec:Mathematical_pre} for a brief introduction, and \cite{Gelb1968} and \cite[Ch. 5]{Slotine1991}  for details. We consider the regime in which a \textit{timescale separation} is present between controller's and crawler's body dynamics, as in \S \ref{sec:GSP}. Within this regime and for the sake of analytical tractability, we introduce the following assumption.

\begin{figure}
    \centering
    \includegraphics[width=\linewidth]{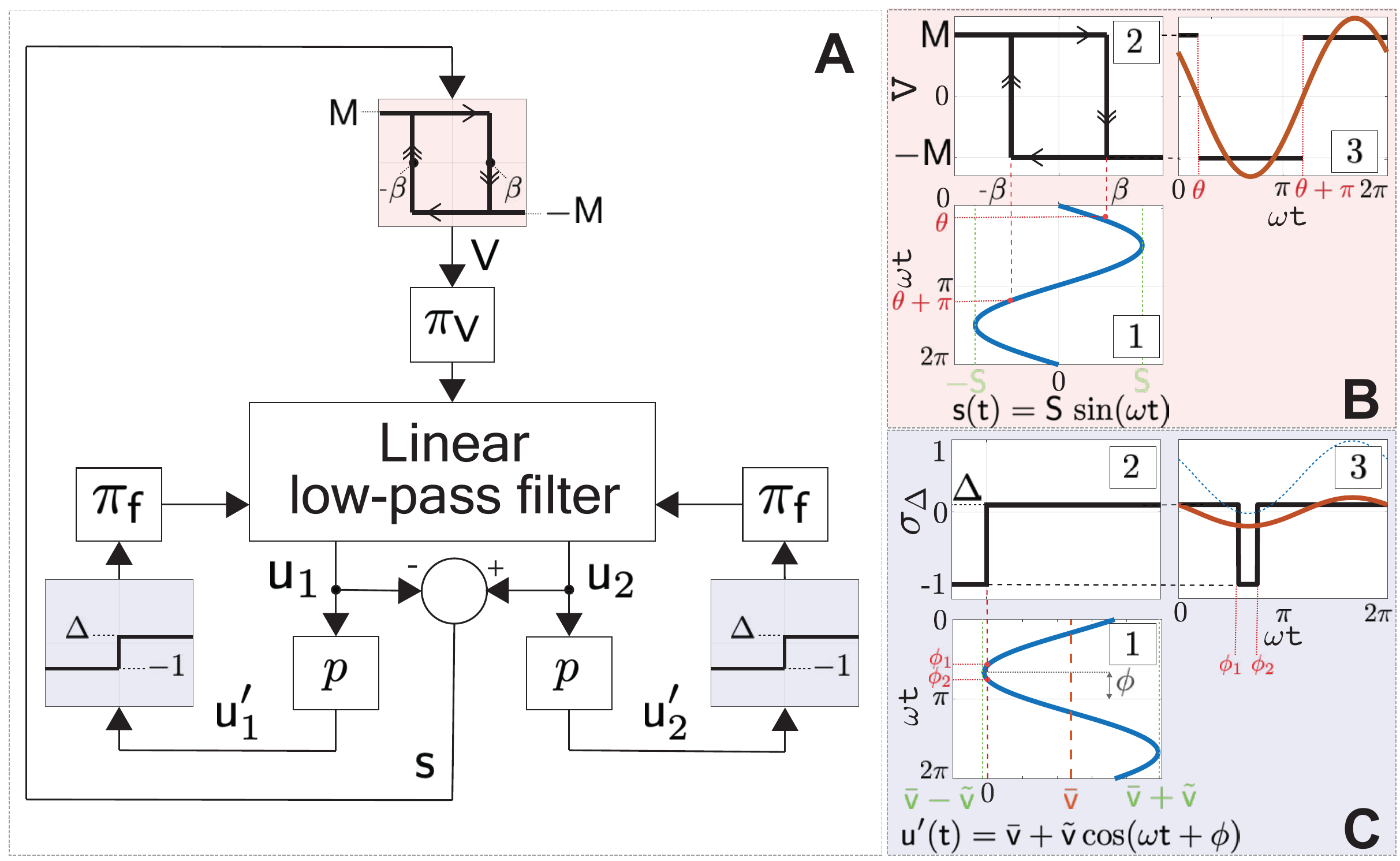}
    \caption{
    \begin{small}
        (\textbf{A}) Block diagram of the singularly-perturbed  excitable crawler dynamics under Assumption \ref{ass:nonlinearities}, splitting linear and nonlinear blocks. 
    The voltage dynamics \eqref{eq:V_dynamics} are replaced by a \textit{static} I/O map---a bi-level hysteretic relay, as shown in the red block. The sigmoidal friction model \eqref{eq:dimensionless_friction} is approximated by an asymmetric relay with parameter $\Delta$, given in \eqref{eq:friction_relay}, shown in the blue block. $p$ denotes the Laplace variable and the $p$-blocks denote time differentiation. \textbf{(B)} Bi-level hysteretic relay relating strain and voltage: (\textbf{1}) sinusoidal strain input $\mathsf{s}$ of amplitude $\mathsf{S}$; (\textbf{2}) bi-level hysteretic relay characteristic, defined by the parameters $\nd{M}$ and $\beta$; (\textbf{3}) voltage output of the relay for the sinusoidal strain input provided in panel (\textbf{1})---in black; its fundamental harmonic is in orange. \textbf{(C)} Asymmetric relay relating local speed and frictional force: (\textbf{1})  speed input $\mathsf{u}'$, with a non-zero average $\bar{\mathsf{v}}$ and harmonic profile, with a phase $\phi$ with respect to the strain;
    (\textbf{2}) asymmetric relay $\sigma_{\Delta}$ relating speed and frictional force; (\textbf{3}) output of the asymmetric relay---in black; its harmonic approximation is in orange and the dotted blue line represents the speed input, showing that input and output are in phase. 
    \end{small}
    }
    \label{fig:harmonic_balance_CL}
\end{figure}

\begin{assumption}[Approximations of nonlinearities]
\label{ass:nonlinearities}
    The following simplifications hold: \textbf{(A)} In the singularly perturbed limit, the relationship between the voltage $\nd{V}$ and strain $\nd{s}$ is approximated by  a \textit{bi-level hysteretic relay}, as depicted in Fig. \ref{fig:harmonic_balance_CL}\textbf{B}. The amplitude of the voltage is denoted by $\nd{M}$ and the switching threshold by $\beta$. The static I/O map is such that $\nd{V}(\nd{s} = \beta) = \nd{M}$ and $\nd{V}(\nd{s} = -\beta) = -\nd{M}$. 
     \textbf{(B)} The smooth friction I/O map \eqref{eq:dimensionless_friction} is approximated by the following piecewise constant model
        \begin{small}
        \begin{equation}
    \sigma_{\Delta}(\nd{u}'):= 
    \begin{cases}
        \Delta, &  \mathsf{u}'\geq 0 \\
        -1, & \mathsf{u}'<0
    \end{cases}
    \label{eq:friction_relay}
    \end{equation}
    \end{small}
where $0< \Delta := \lim_{\nd{u}' \to \infty} \sigma(\nd{u}') = \frac{1 - \tanh(\nd{n_f})}{1+\tanh(\nd{n_f)}} \lll 1$.  A graphical representation of \eqref{eq:friction_relay} is provided in Fig. \ref{fig:harmonic_balance_CL}\textbf{C}.
\end{assumption}

The main result in this section is to prove that optimal crawling corresponds to a \textit{resonant} limit cycle in the system---that is, the optimal frequency of the self-sustained oscillations in the excitable crawler is the natural frequency of the crawler's body, $\omega_n$, as given in \eqref{eq:natural_freq}.  This resonance is facilitated by a \textit{neuromechanical matching condition} \cite{Arbelaiz2025}, in which the switching threshold of the voltage ($\beta$) matches the amplitude of the strain in the soft crawler's body ($\nd{S}$). When such a matching condition is met, there is a \textit{phase alignment} between the voltage produced by the controller and the strain rate in the crawler's body. In turn, this alignment implies the non-negativity of the instantaneous power provided by the actuator to the crawler's body and the maximization of the corresponding average power. The remainder of this section deals with the formalization of such claims. \S \ref{subsec:DF_and_HB} introduces preliminary derivations---the fundamental harmonic approximations to the output of the nonlinearities in the excitable crawler and the  harmonic balance equations---needed to prove the main result. \S \ref{subsec:com_speed_max} formulates the optimization problem to maximize the average speed of the crawler's center of mass, based on the results from the harmonic balance analysis. \S \ref{subsec:resonance_and_matching} solves the optimization problem analytically and presents physical insights into its optimal solution.

Although our resonance optimality result is derived for the two-segment crawler morphology, it is consistent with the broader empirical observation that locomotion performance in soft and vibration-driven robots is often enhanced when actuation is tuned to the mechanical response of the body. Thus, our analysis suggests a more general approach to principled locomotion design, extendable to other morphologies, including undulatory, swimming, pulse-jet, and vibration-driven robots.

\vspace{-.25cm}
\subsection{Harmonic approximations \& harmonic balance}
\label{subsec:DF_and_HB}
\vspace{-0.1cm}
We start by providing the fundamental harmonic approximations to the outputs of the nonlinear blocks in Fig. \ref{fig:harmonic_balance_CL}. 

\subsubsection{Fundamental harmonic of the frictional force}
 Let 
 \begin{small}
\begin{equation}
  \mathsf{u}'(\nd{t}) = \bar{\nd{v}} + \tilde{\nd{v}} \cos(\omega \nd{t} + \phi) 
    \label{eq:HB_friction_input}
\end{equation}  
 \end{small}
with
$
  \bar{\nd{v}} >0 \text{ and } \tilde{\nd{v}}>0 
$
and define 
$
    \nd{a}:= \frac{\bar{\nd{v}}}{\tilde{\nd{v}}}>0.
$
The frictional force is approximated by its fundamental component when subject to the input \eqref{eq:HB_friction_input}:
\begin{small}
    \begin{align}
        \sigma_\Delta(\nd{u}') &  = \sigma_\Delta \big( \bar{\nd{v}} + \tilde{\nd{v}} \cos(\omega \nd{t} + \phi) \big) \nonumber \\
        & \approx \frac{\sigma_0}{2} + \sigma_{\nd{cos}} \cos(\omega \nd{t}) + \sigma_{\nd{sin}} \sin(\omega \nd{t}), 
    \end{align}
\end{small}
\text{where}
\begin{small}
$
    \sigma_0  := \frac{2}{\pi} \Big( \pi \Delta - \arccos(\nd{a})(\Delta + 1)\Big), \label{eq:friction_DC}\; 
    \sigma_{\nd{sin}}  = - \frac{2}{\pi} (1+\Delta) \sin\phi \sqrt{1-\nd{a}^2}, \label{eq:friction_sin} \text{ and }
    \sigma_{\nd{cos}} 
     = \frac{2}{\pi} (1+\Delta) \cos\phi \sqrt{1-\nd{a}^2}. \label{eq:friction_cos}
\label{eq:HB_friction_coeffs_summary}
$
\end{small}
\label{eq:HB_friction_coeffs}
The detailed computation of the coefficients is provided in \cite[Appendix A]{Shen2024}.
The frictional force and its fundamental harmonic are depicted in Fig. \hyperlink{fig:harmonic_balance_CL}{5\textbf{C}}.

\subsubsection{Fundamental harmonic of the voltage}
Let the strain be a sinusoid of the form 
\begin{small}
\begin{equation}
    \nd{s}(\nd{t}) = \nd{S} \sin(\omega \nd{t}),
    \label{eq:strain_FH_approx}
\end{equation}
\end{small}
where the following assumption holds
\begin{assumption}
    \label{assumption:s_beta}
     $\nd{S} \geq \beta > 0$ and $\omega > 0$, where $\beta$ is the dimensionless switching threshold of Assumption~\ref{ass:nonlinearities}. 
\end{assumption}
Assumption \ref{assumption:s_beta} ensures that the amplitude of the strain in the crawler's body is sufficiently large for the voltage $\nd{V}$ to switch signs according to the bi-level hysteretic relay approximation.
The square wave voltage output is approximated by its fundamental harmonic component as
\begin{small}
    \begin{align}
        \nd{V}(\nd{s})
        & \approx \nd{V}_0 + \nd{V}_{\nd{sin}} \sin(\omega \nd{t}) + \nd{V}_{\nd{cos}} \cos(\omega \nd{t}),
    \end{align}
\end{small}
    with
    $
        \nd{V}_0 = 0,\; 
        \nd{V}_{\nd{sin}}  = -\frac{4}{\pi} \, \nd{M}\,  \sqrt{1 - \Big(\frac{\beta}{\nd{S}}\Big)^2}, \text{ and }
        \nd{V}_{\nd{cos}}  = \frac{4}{\pi} \, \nd{M}\,  \frac{\beta}{\nd{S}},
    $
where the bi-level relay parameters $\nd{M} \in \mathbb{R}_{++}$ and $\beta \in \mathbb{R}_{++}$ are as shown in Fig. \hyperlink{fig:harmonic_balance_CL}{5\textbf{B}}. The detailed derivation of the coefficients is provided in Appendix \hyperlink{subsec:fundamental_H_voltage}{III.A}.

\begin{rem}
    The relative phase between input (strain) and output (voltage) to the bi-level hysteretic relay depicted in Fig. \hyperlink{fig:harmonic_balance_CL}{5\textbf{B}} is determined by the ratio $\beta/\nd{S}$. 
    This ratio plays an important role in the efficiency of the gait of the excitable crawler, as it determines 
    the sign of the instantaneous power that the actuator provides to the soft robot. 
\end{rem}

\subsubsection{Harmonic balance}
\label{subsubsec:harmonic_balance} 

Through harmonic balance analysis, we aim to obtain the dependence of the average speed of the crawler's center of mass, denoted $\bar{\nd{v}}_{\nd{com}}$, on the mechanical and electrical parameters. Ultimately, our goal is to maximize $\bar{\nd{v}}_{\nd{com}}$. We approximate the signals by their respective fundamental harmonics. We ground the phase of the strain $\nd{s}$ and define the phases of the remaining signals relative to that of the strain:
\begin{small}
\begin{equation}
        \nd{s}(\nd{t})  = \nd{S} \sin(\omega \nd{t}), \; 
        \nd{s}'(\nd{t}) = \omega \nd{S} \cos(\omega \nd{t}), \; 
         \nd{s}''(\nd{t}) = -\omega^2 \nd{S} \sin(\omega \nd{t}).
    \label{eq:strain_harmonics}
\end{equation}
\end{small}
The segmental speeds and accelerations are approximated by
\vspace{-0.1cm}
\begin{small}
\begin{subequations}
    \begin{align}
    \nd{u}_\nd{i}'(\nd{t}) & = \bar{\nd{v}}_\nd{i} +\tilde{\nd{v}}_\nd{i} \cos(\omega \nd{t} + \phi_\nd{i}), \label{eq:speed1_harmonic_approx}\\
    \nd{u}_\nd{i}''(\nd{t}) & = - \omega \tilde{\nd{v}}_\nd{i} \sin(\omega \nd{t} + \phi_\nd{i}), \label{eq:acc1_harmonic_approx} 
    \end{align}
\end{subequations} 
\end{small}
where $\nd{i} \in \{1,2\}$ and we let the following assumption hold:
\begin{assumption}
\label{assumption:positive_amplitudes}
    W.l.g., $\bar{\nd{v}}_\nd{i}, \tilde{\nd{v}}_\nd{i} >0, \; \nd{i}\in \{1,2\}.$
\end{assumption}
From the previous analysis, 
\begin{small}
\begin{equation}
    \nd{V}(\nd{t}) \approx \frac{4 \nd{M}}{\pi} \bigg( -\sqrt{1 - \Big(\frac{\beta}{\nd{S}}\Big)^2} \sin(\omega \nd{t}) + \frac{\beta}{\nd{S}} \cos(\omega \nd{t})\bigg).
    \label{eq:voltage_fundamental_component}
\end{equation}
\end{small}
We aim to solve for $\{\omega, \nd{S}, \bar{\nd{v}}_1, \tilde{\nd{v}}_1, \bar{\nd{v}}_2, \tilde{\nd{v}}_2, \phi_1, \phi_2\}$ parametrically, as a function of the dimensionless numbers in the excitable crawler dynamics. Harmonic balance\footnote{To streamline the presentation, technical details of the harmonic balance analysis have been deferred to Appendix \hyperlink{subsec:harmonic_balance}{III.B}.} yields $\phi_1 = \pi, \phi_2 = 0, \nd{a}:= \nd{a}_1 = \nd{a}_2 = \cos\Big( \frac{\pi \Delta}{1 + \Delta}\Big), \bar{\nd{v}}:= \bar{\nd{v}}_1 = \bar{\nd{v}}_2 = \nd{a} \tilde{\nd{v}}$, where $\tilde{\nd{v}} := \tilde{\nd{v}}_1 = \tilde{\nd{v}}_2$. The following identities
close the system:
\begin{small}
\begin{subequations}
    \begin{align}
        \tilde{\nd{v}} & = \frac{\omega \nd{S}}{2}, \label{eq:balance_final_1}\\
         0  & = -2 \zeta \nd{S} \omega + \frac{8 \pi_\nd{V}}{\pi} \frac{\nd{M} \beta}{\nd{S}} - \frac{4 \pi_\nd{f}}{\pi} (1+ \Delta) \sqrt{1-\nd{a}^2},
        \label{eq:balance_final_2} \\
        0 & = \frac{8 \pi_\nd{V} \nd{M}}{\pi} \sqrt{1 - \Big( \frac{\beta}{\nd{S}}\Big)^2} + \nd{S} (1-\omega^2). 
        \label{eq:balance_final_3} 
    \end{align}
    \label{eq:balance_closure}
\end{subequations}
\end{small}
There is no need to explicitly solve  \eqref{eq:balance_closure} to perform the optimization we seek.

\subsection{Optimization of the bi-level relay: 
maximizing  the soft crawler's speed}
\label{subsec:com_speed_max}
We formulate an optimization problem to find the value of the parameter $\beta$ maximizing the average speed of the crawler's center of mass, $\bar{\nd{v}}_{\nd{com}}$, given by
$
    \bar{\nd{v}}_\nd{com} = \frac{\bar{\nd{v}}_1 + \bar{\nd{v}}_2}{2} = \bar{\nd{v}} = \nd{a}(\Delta) \tilde{\nd{v}} = \nd{a}(\Delta) \frac{\omega \nd{S}}{2}.
    \label{eq:v_com}
$
Since $\nd{a}$ only depends on the friction asymmetry $\Delta$,
$
    \mathrm{arg} \max_{\beta \in \mathbb{R}_{++}} \; \bar{\nd{v}}_\nd{com} = \mathrm{arg}  \max_{\beta \in \mathbb{R}_{++}} \; \omega \nd{S}.
$
From \eqref{eq:balance_final_2}, 
\begin{small}
$
    \omega(\nd{S}) 
    = \frac{1}{2 \zeta \nd{S}} \Big( \alpha \frac{\beta}{\nd{S}} - \eta \Big),
    \label{eq:omega_in_S}
$
\end{small}
where we have defined the constants 
\begin{small}
\begin{align}
    \alpha := \frac{8 \pi_\nd{V} \nd{M}}{\pi}  \text{ and }
    \eta  := 
    \frac{4 \pi_\nd{f}}{\pi} (1+ \Delta) \sin\bigg( \frac{\pi\Delta}{1+\Delta}\bigg). 
    \label{eq:alpha_beta_const}
\end{align}
\end{small}
\begin{assumption}[$\alpha > \eta$]
\label{ass:alpha_greater_eta}
    From this point onward we assume that the parameter $\eta$ and $\alpha$ defined in \eqref{eq:alpha_beta_const} satisfy
    $
        \alpha > \eta.
    $ 
\end{assumption}

Due to the coupling in the harmonic balance equations, the $\bar{\nd{v}}_\nd{com}$ maximization problem is formulated as a joint optimization in the decision variables $(\beta, \nd{S})$. 
Introducing the change of variable $\nd{Z} := \frac{\beta}{\nd{S}}$ the problem reads 
\begin{subequations}
\begin{small}
\begin{maxi}|s|[0]                    
{(\nd{S}, \nd{Z}) \in \mathbb{R}^2}
{
\begin{aligned}
     {\nd{Z}}
\end{aligned}
} 
{\label{optimizationProblem2}} 
{} 
\addConstraint{ 
\nd{g}(\nd{S}, \nd{Z})}{=0} 
\addConstraint{\nd{S}}{>0} 
\addConstraint{\frac{\eta}{\alpha}<\nd{Z}}{\leq  1,} 
\end{maxi}
\end{small}
where $\alpha$ and $\eta$ are as defined in \eqref{eq:alpha_beta_const}, and 
\begin{small}
  \begin{equation}
        \nd{g}(\nd{S}, \nd{Z}):= \alpha \sqrt{1 - \nd{Z}^2} + \nd{S} \Big( 1 - \frac{1}{4 \zeta^2 \nd{S}^2}( \alpha \nd{Z} - \eta)^2\Big).
        \label{eq:g_def}
    \end{equation}
\end{small}
\label{eq:HB_opt}
\end{subequations}
By Assumption \ref{assumption:s_beta}, $0<\nd{Z}\leq 1$. Moreover, since $\omega>0 \, \Rightarrow \, \nd{Z}>\frac{\eta}{\alpha}$. Together, they define the bounds for $\nd{Z}$ in \eqref{eq:HB_opt}. The optimal solution $(\nd{S}^*, \nd{Z}^*)$ to \eqref{eq:HB_opt} is to select the largest $\nd{Z} \in (\frac{\eta}{\alpha},1]$, such that the value  $\nd{S}^*$ satisfying $\nd{g}(\nd{S}^*, \nd{Z}^*) = 0$ is $\nd{S}^*>0$. Lemma \ref{lemma:nonempty_set} establishes the feasibility of the problem.

\begin{lem}[The feasible set of \eqref{eq:HB_opt}]
\label{lemma:nonempty_set}
Consider the optimization problem \eqref{eq:HB_opt} with $\alpha$ and $\eta$ as defined in \eqref{eq:alpha_beta_const}
and satisfying Assumption \ref{ass:alpha_greater_eta}.
 Let $\nd{g}: \mathbb{R}_{++} \times (0,1] \to \mathbb{R}$ be as defined in \eqref{eq:g_def}. Then,
 
        a) The feasible set of \eqref{eq:HB_opt}   is nonempty.
        
        b) For a fixed value $\nd{Z} \in (\frac{\eta}{\alpha},1] $, there is a unique value $\nd{S}_0 \in [0,\infty)$ such that $\nd{g}(\nd{S}_0,\nd{Z}) = 0$.
\end{lem}

\begin{proof}
$\nd{g}(\nd{S}; \nd{Z})$ can be viewed as a univariate function of $\nd{S}$ parametrized by $\nd{Z}\in (\frac{\eta}{\alpha},1]$.  Then,
    \textit{a)} $\lim_{\nd{S} \to \infty} \nd{g}(\nd{S}; \nd{Z}) = \infty$ and $\lim_{\nd{S} \to 0^+} \nd{g}(\nd{S}; \nd{Z}) = - \infty$. 
    Furthermore, since $\nd{g}$ is continuous in its domain, 
    by the Intermediate Value Theorem there exists $\nd{S}_0 \in (0,\infty)$ such that $\nd{g}(\nd{S}_0; \nd{Z}) = 0 \, \Rightarrow $ the feasible set of \eqref{optimizationProblem2} is non-empty. \textit{b)} $\partial_\nd{S} \nd{g}(\nd{S}; \nd{Z}) = 1 + \frac{1}{4 \zeta^2 \nd{S}^2}(\alpha \nd{Z} - \eta)^2  > 0$; together with the properties derived in part \textit{a)}
    $\Rightarrow$ $\nd{S}_0 \in (0, \infty)$ is unique.
\end{proof}
Lemma \ref{lemma:nonempty_set} establishes two properties of the constrained optimization problem \eqref{eq:HB_opt}: the first is that it is feasible; the second is that for any given value of $\nd{Z} \in (\frac{\eta}{\alpha},1]$  there exists a unique $\nd{S}_0 \in [0,\infty)$ in the feasible set. Together, these two properties imply that the unique maximizer is $\nd{Z}^* = 1$.

\subsection{Crawling at resonance is optimal and enabled by a neuromechanical matching condition}
\label{subsec:resonance_and_matching}
\begin{thm}[The optimal solution to \eqref{eq:HB_opt}]
\label{thm:optimal_solution}    
Consider the constrained optimization \eqref{eq:HB_opt} with $\alpha$ and $\eta$ as defined in \eqref{eq:alpha_beta_const} and satisfying Assumption \ref{ass:alpha_greater_eta}. Its unique maximizer is given by
\begin{equation}
\nd{Z}^*= 1 \text{ and } \nd{S}^* = \frac{1}{2 \zeta}(\alpha - \eta).
    \label{eq:maximizer}
\end{equation}
\end{thm}
\begin{proof}
    Within $\nd{Z}\in (\frac{\eta}{\alpha},1], \, \nd{Z}^* = 1$ maximizes the cost function in \eqref{optimizationProblem2}. It remains to be proved that it produces a feasible value of $\nd{S}$ and that such value is unique. This is guaranteed by Lemma \ref{lemma:nonempty_set}.
    Substitution of $\nd{Z}^* = 1$ in \eqref{eq:g_def} yields
    \eqref{eq:maximizer}, which satisfies $\nd{S}>0$ by Assumption \ref{ass:alpha_greater_eta}.
\end{proof}

\begin{figure}
    \centering
    \includegraphics[width=0.49\textwidth]{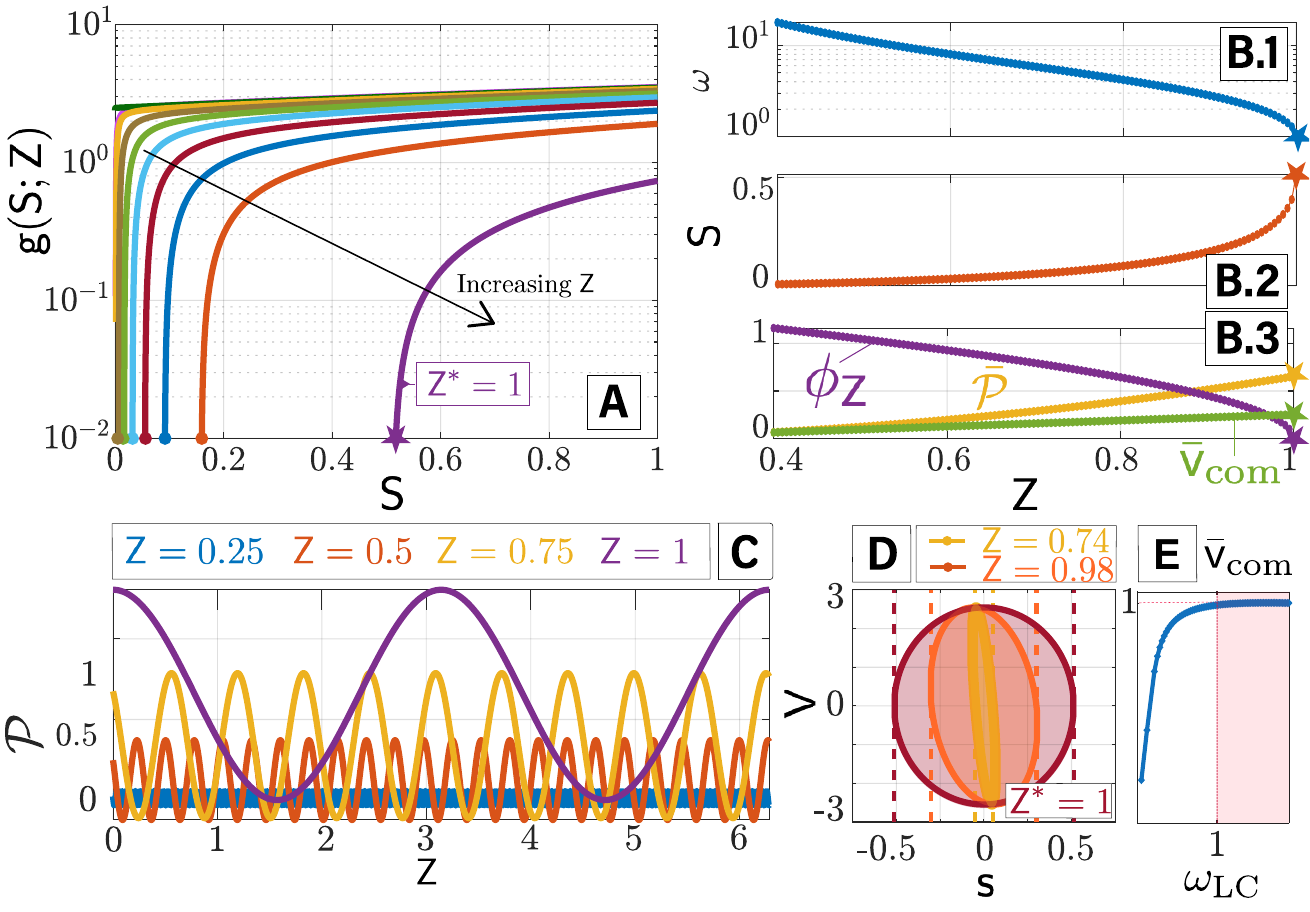}
    \caption{\begin{small}
    Describing function analysis and optimization. Parameter values: $\zeta = 2, \pi_\nd{V} = 0.5, \nd{M} = 2, \pi_\nd{f} = 2.5, \nd{n_f} = 1.5$.  \textbf{(A)} $\nd{g}$ as defined in \eqref{eq:g_def} as a function of $\nd{S}$ for $\nd{Z} \in \{0.1; 0.2; 0.3; 0.4; 0.5; 0.6; 0.7; 0.8; 0.9; 1 \}$, as  indicated. Vertical axis is in logarithmic scale. Per Lemma \ref{lemma:nonempty_set}, each curve has a unique intersection with the abscissa---respectively highlighted by a circle.  The optimal $\nd{S}$ as given in \eqref{eq:maximizer} which corresponds to $\nd{Z}^* = 1$ is highlighted by a star. For the parameter values selected in this example, $\nd{S}^* \approx 0.51$. \textbf{(B)} Quantities of interest as a function of $\nd{Z}$, as approximated by the describing function and harmonic balance analysis: \textbf{(B.1)} the (dimensionless) frequency of the state oscillations, $\omega$. The optimal $\omega^* = 1$ corresponds to crawling at resonance; \textbf{(B.2)} amplitude $\nd{S}$ of the strain in the soft crawler's body; \textbf{(B.3)} (purple) relative phase between the fundamental harmonics of voltage $\nd{V}$ and the strain rate $\nd{s}'$. The fundamental harmonic approximations of these two signals align (i.e., $\phi_\nd{Z} = 0$) at the optimum $\nd{Z}^* = 1$; (green) average speed of the excitable crawler's center of mass, $\bar{\nd{v}}_{\nd{com}}$; (yellow) average actuator power $\bar{\mathcal{P}}$, maximized at $\nd{Z}^*=1$. \textbf{(C)} Instantaneous actuator power $\mathcal{P}$ as a function of time for different values of $\nd{Z}$ as highlighted in the legend. $\mathcal{P}$ oscillates at $2 \omega$. $\mathcal{P}$ is a \textit{non-negative} signal only at optimality ($\nd{Z}^*=1$, purple curve). 
    \textbf{(D)} Under periodic motion the actuator force versus strain trajectories form a closed curve.  The work over a cycle done by the actuator force on the crawler's segments---for different values of $\nd{Z}:= \beta/\nd{S}$, as indicated---is the area enclosed clockwise by the respective curve. 
    Dotted lines represent the respective values of $\beta$. The actuator's force work is maximized at $\nd{Z}^* = 1$.
    As such matching condition, the amplitude of the strain is maximal and the closed curve tangent to the vertical at $\nd{S} = \beta$, indicating maximal actuator work per cycle.  \textbf{(E)} Full smooth-model validation. Normalized average center-of-mass speed of the attracting limit cycle as a function of its frequency \(\omega_{\rm LC}\), obtained by sweeping $\pi_\nd{s}$ in the original smooth model \eqref{eq:dimensionless_electromechanical_dynamics_movingFrame}. The speed rises sharply as \(\omega_{\rm LC}\) approaches \(1\), corresponding to the crawler body’s natural frequency in dimensional coordinates, and then enters a near-maximal plateau region (shaded in red).
     \end{small}
    }
\label{fig:harmonic_balance_opt}
\end{figure}

Fig. \ref{fig:harmonic_balance_opt}\textbf{A} illustrates the results of Lemma \ref{lemma:nonempty_set} and Theorem \ref{thm:optimal_solution}. 
In terms of the parameters in the dynamics, the optimal solution \eqref{eq:maximizer} is 
\begin{subequations}
\begin{align}
    \beta^* & = \nd{S}^*, \;\; \text{(neuromechanical matching condition)} \label{eq:opt_beta} \\
    \nd{S}^* & = \underbrace{\frac{2}{\pi \zeta}}_{\substack{\text{body viscosity/} \\ \text{damping}}} \Big(   \underbrace{2 \pi_\nd{V} \nd{M} \vphantom{\frac{2}{\pi \zeta}}}_{\substack{\text{electric voltage/}\\ \text{neuron}}}  -\underbrace{\pi_\nd{f} \, (1+\Delta) \sin \Big( \frac{\pi \Delta}{1+\Delta }\Big) \vphantom{\frac{2}{\pi \zeta}}}_{ \substack{\text{frictional force/} \\ \text{environment}} }   \Big).
    \label{eq:opt_strain_ampl}
\end{align}
\label{eq:opt_sol}
\end{subequations}
The corresponding maximal average speed of the crawler's center of mass is
    \begin{small}
    \begin{equation*}
    \bar{\nd{v}}_{\nd{com}}^* = \frac{1}{\pi \zeta} \cos \bigg( \frac{\pi \Delta}{1 + \Delta}\bigg) \Big( 2 \pi_\nd{V} \nd{M}  - \pi_\nd{f} (1 + \Delta) \sin\bigg( \frac{\pi \Delta}{1 + \Delta} \bigg)\Big).
    \label{eq:v_com_max}
    \end{equation*}
    \end{small}
\eqref{eq:opt_beta} reveals that optimal crawling gaits are enabled by \textit{a matching between electrical (neuronal) and mechanical (body) scales}. In particular, the optimal value of $\beta$---which defines the threshold in strain for the voltage to switch its polarity---matches the optimal amplitude of the strain in crawler's body. We refer to \eqref{eq:opt_beta} as a \textit{neuromechanical matching condition.} Such an \textit{optimal matching condition} is alternatively formulated in terms of a resonance and non-negative instantaneous actuation power in Corollary \ref{cor:resonance_optimal}.  \eqref{eq:opt_strain_ampl} shows how the amplitude of the strain of the soft-bodied crawler responds to external and internal forces.
Accordingly, in a controller-plant co-design process the values of some of these parameters can be selected to ensure safe strain levels in the crawler during operation. Alternatively, the optimization  \eqref{optimizationProblem2} can be modified to incorporate constraints on the strain amplitude.

\begin{cor}[The neuromechanical matching condition corresponds to crawling at resonance and maximizes actuation power]
\label{cor:resonance_optimal}
The relay parameter $\beta^*$ and amplitude of the crawler's body strain $\nd{S}^*$ as given in \eqref{eq:opt_sol}---optimal solution to the optimization problem \eqref{eq:HB_opt}:

     a) correspond to crawling at resonance. That is, the optimal (dimensional) forcing frequency is $\omega_n$---the undamped natural frequency of the linearized crawler \eqref{eq:natural_freq}; 
     
    b)  maximize the average actuator power over a crawling cycle and make the instantaneous actuator power non-negative. These effects are due to the phase alignment between the strain rate in the soft crawler's body and the voltage at the neuromechanical matching condition. 
\end{cor}
\begin{proof} 
   a) To show that crawling at resonance is optimal, substitute the optimal solution \eqref{eq:opt_sol} into the harmonic balance expression for $\omega(\nd{S})$, which provides an  optimal \textit{dimensionless} frequency of $\omega^* = 1$.  Taking into account that $\omega = t_* \cdot \Omega$, where $\Omega$ denotes the dimensional frequency and $t_*$ is as defined in \eqref{eq:mechanical_scales}, we obtain that $\Omega^* = \frac{\omega^*}{t_*} = \sqrt{\frac{2k}{m}} = \omega_n$.\\
b) We define the instantaneous actuator power as
\begin{small}   
$
 \mathcal{P}(\nd{t}) := \nd{s}'(\nd{t}) \nd{V}(\nd{t}) = \omega \nd{S} \cos(\omega \nd{t})\big( \nd{V}_{\sin} \sin(\omega \nd{t}) + \nd{V}_{\cos} \cos(\omega \nd{t})\big)  = \omega \nd{S} \nd{V}_{\sin} \frac{\sin(2 \omega \nd{t})}{2} + \omega \nd{S} \nd{V}_{\cos} \frac{\cos(2 \omega \nd{t}) + 1}{2},
$
\end{small}
which oscillates at a \textit{higher harmonic} than the state. Accordingly, in steady-state, the average actuator power\footnote{Since both $\nd{s}'$ and $\nd{V}$ are periodic signals with the same period, so is their product. Thus, the definition of the average power $\bar{\mathcal{P}}$ is \textit{independent} of $t$. We highlight such independence through the limits of integration in the definition of $\bar{\mathcal{P}}$.} over a cycle is
$
    \bar{\mathcal{P}} :=  \frac{1}{T_{\mathcal{P}}} \int_t^{T_{\mathcal{P}}+t} \mathcal{P}(\nd{t}) d \nd{t}  = \frac{\omega \nd{S}}{T_{\mathcal{P}}} \int_t^{T_{\mathcal{P}}+t} \big(  \nd{V}_{\sin} \frac{\sin(2 \omega \nd{t})}{2} + \nd{V}_{\cos} \frac{\cos(2 \omega \nd{t}) + 1}{2} \big) d\nd{t}  = \frac{\omega \nd{S}}{2} \nd{V}_{\cos} = \frac{\nd{M}}{\zeta \pi} \nd{Z} (\alpha \nd{Z} - \eta),
$
where $T_{\mathcal{P}} = \frac{\pi}{\omega}$ denotes the period of the instantaneous actuation  power, $\alpha$ and $\eta$ are as defined in \eqref{eq:alpha_beta_const}, and $\nd{Z}:= \beta/\nd{S}$. Thus, the average power $\bar{\mathcal{P}}$
is only a function of $\nd{Z}$.  Within the domain $\nd{Z}\in (\frac{\eta}{\alpha},1]$, $\bar{\mathcal{P}}$ is maximized at $\nd{Z}^* = 1$. At this neuromechanical matching, the fundamental harmonic of the voltage signal consists only of the cosine component (i.e.,  $\nd{V}_{\sin} = 0$ and $\nd{V}_{\cos} = \frac{4\nd{M}}{\pi}$ is maximized). Consequently, the relative phase $\nd{\phi}_{\nd{Z}}$ between the voltage and the strain rate is $\nd{\phi}_{\nd{Z}^*} = 0$. Such a phase alignment makes the instantaneous actuator power a non-negative function $\mathcal{P}(\nd{t})  =    \frac{8  \nd{M}}{\pi^2 \zeta} \big( 2 \pi_\nd{V} \nd{M}  - \pi_\nd{f} \, (1+\Delta) \sin(\frac{\pi \Delta}{1+\Delta}) \vphantom{\frac{2}{\pi \zeta}}  \big)  \cos^2(\nd{t}) \geq 0, \, \forall \nd{t}$.
\hfill
\end{proof}

Fig. \ref{fig:harmonic_balance_opt}\textbf{B} shows different quantities of interest---approximated in steady-state through the respective fundamental harmonics---in the excitable crawler as a function of $\nd{Z}:= \beta/\nd{S}$. 
  Higher values of $\nd{Z}$ provide higher strain amplitudes and lower oscillation frequencies, that is, longer crawling gaits at higher strain levels. These provide faster crawling speeds (see the green curve representing $\bar{\nd{v}}_{\nd{com}}$ in Fig. \ref{fig:harmonic_balance_opt}\textbf{B}.3) and a higher average actuator power over a crawling cycle (see the yellow curve  representing $\bar{\mathcal{P}}$ in Fig. \ref{fig:harmonic_balance_opt}\textbf{B}.3). Fig. \ref{fig:harmonic_balance_opt}\textbf{B} also shows the alignment between $\nd{s}'$ and $\nd{V}$ at optimality (i.e., a relative phase of $\nd{\phi}_{\nd{Z}^*} = 0$). Such an alignment results in a non-negative instantaneous actuator power---illustrated in Fig. \ref{fig:harmonic_balance_opt}\textbf{C}---characteristic of dynamical systems with periodic harmonic steady-state motion operating at resonance.
The event-timing of the actuator in this case produces the largest energy transfer from the actuator to the segments of the crawler---so that actuator effort reinforces the relevant mechanical mode and contributes to center-of-mass motion. Such work maximization at $\nd{Z}^* = 1$ is illustrated in Fig. \ref{fig:harmonic_balance_opt}\textbf{D}. For arbitrary values of $\nd{Z}$, the steady-state instantaneous actuation power might become negative over some time intervals (as illustrated in Fig. \ref{fig:harmonic_balance_opt}\textbf{C}).

\section{Conclusion \& Perspectives}
\label{sec:conclusion}

\textit{Soft robots} are poised to revolutionize diverse industries and applications due to their inherent flexibility and adaptability. These properties facilitate successful interactions with uncertain, dynamic, and unstructured environments, while offering improved energy efficiency and robustness. 
Yet their broader deployment remains limited by the lack of principled control-synthesis frameworks that explicitly exploit the distinctive features of soft bodies and their nonlinear dynamics.

In this work, we propose \textit{spiking control systems} as a paradigm for soft robotic control and analyze an \textit{excitable controller} driving locomotion in a single-input single-output soft robotic crawler. Through bifurcation analysis, we characterize how the sensorimotor gain shapes equilibrium stability and the emergence of peristaltic waves. This gain 
can be endowed with dynamics from state feedback or external inputs to enhance control flexibility. Dimensional analysis reveals a separation between electrical and mechanical timescales, which we leverage through Geometric Singular Perturbation Theory to characterize the resulting oscillations.   Finally, we formulate and solve an optimization problem in the singularly perturbed regime, obtaining controller parameters that maximize locomotion speed and showing that optimal crawling occurs at mechanical resonance.

Together, our analysis tools and results provide a control-theoretic foundation for designing and analyzing spiking controllers in soft robotics, with potential to extend to more complex and distributed morphologies. Natural directions for future work include  implementation and experimental evaluation of the proposed controller on compliant robotic platforms such as the origami-enabled multisegment crawler studied in \cite{Earnhardt2025},  extensions to multi-segmented robotic architectures, and adaptive regulation of electrical timescales to match mechanical scales for further efficiency—paralleling homeostatic regulation observed in biological neural circuits.

\section*{Acknowledgements}

The first author acknowledges useful discussions on dimensionality reduction and the numerical continuation software \textsc{MatCont} 
with Prof. Anastasia Sergeyevna Bizyaeva (Cornell U.), as well as discussions with Ph.D. students Ian Xul Belaustegui and Marcela Ordorica (Princeton U.).

\section*{Appendix I: Proofs of Section \ref{sec:organizing_bifurcations}}
\label{sec:Appendix_II}

\subsection{Preliminaries: Derivatives of $\boldsymbol{f}$}

The analysis of bifurcations in the excitable crawler dynamics \eqref{eq:dimensionless_electromechanical_dynamics_movingFrame} requires the computation of directional derivatives of several orders of the vector field $\boldsymbol{f}$. We provide the derivatives needed for some subsequent calculations.

\subsubsection{First order derivatives}
\label{subsubsec:first_order_ders}

The Jacobian of the vector field $\boldsymbol{f}$ in \eqref{eq:dimensionless_electromechanical_dynamics_movingFrame} is:
\begin{small}
\begin{equation}
    \nabla_{\boldsymbol{\nd{x}}} \boldsymbol{f}(\boldsymbol{\nd{x}}) = 
    \begin{bmatrix}
    -3 \pi_{\nd{c}} \nd{V}^2 + \pi_{\nd{l}} & 0 & - \pi_{\nd{s}} &  0\\
    0 & -\frac{\pi_{\nd{f}}}{2} \delta_+ & 0 & -\frac{\pi_{\nd{f}}}{4} \delta_-\\
    0 & 0 & 0 & 1\\
    2 \pi_{\nd{V}} & -\pi_{\nd{f}}\delta_- & - 1 & -\frac{\pi_{\nd{f}}}{2}\delta_+ - 2 \zeta
    \end{bmatrix},
    \label{eq:jacobian_general}
\end{equation}
\end{small}
where we defined
\begin{small}
$
    \delta_{\pm}  := \sigma_{\pi}^{\prime}(\nd{v}_{\text{com}} + \frac{\nd{v_s}}{2}) \pm \sigma_{\pi}^{\prime}(\nd{v}_{\text{com}} - \frac{\nd{v_s}}{2}),\\
$
\end{small}
and denoted
$\frac{d \sigma_\pi}{dx} = \sigma_{\pi}^{\prime}(x) = \frac{\pi_{\varepsilon}\text{sech}^2(\pi_{\varepsilon} x + \nd{n_f})}{1+\tanh{\nd{n_f}}}$.

\subsubsection{Second order derivatives} 
\label{subsubsec:second_order_ders} 
We provide those second order derivatives of $\boldsymbol{f}$ that do not vanish identically and avoid explicitly stating symmetric derivatives (Schwarz's theorem)
\begin{subequations}
\begin{small}
    $
        \frac{\partial^2 \boldsymbol{f}}{\partial \nd{V}^2}(\boldsymbol{\nd{x}})   = \big[ - 6 \pi_{\nd{c}} \nd{V}\;\; 0 \;\; 0 \;\; 0\big]^\top, 
        \;
         \frac{\partial^2 \boldsymbol{f}}{\partial \nd{v_{com}}^2}(\boldsymbol{\nd{x}})   = \big[ 0\;\; -\frac{\pi_{\nd{f}}}{2} \delta_+' \;\; 0 \;\; -\pi_{\nd{f}} \delta_-' \big]^\top, \; 
         \frac{\partial^2 \boldsymbol{f}}{\partial \nd{v_s}^2}(\boldsymbol{\nd{x}})   = \big[ 0\;\; -\frac{\pi_{\nd{f}}}{8} \delta_+' \;\; 0 \;\; -\frac{\pi_{\nd{f}}}{4} \delta_-'\big]^\top,\; 
        \frac{\partial^2 \boldsymbol{f}}{\partial \nd{v_s }\partial \nd{v_{com}}}(\boldsymbol{\nd{x}})    = \big[ 0\;\; -\frac{\pi_{\nd{f}}}{4} \delta_-' \;\; 0 \;\; -\frac{\pi_{\nd{f}}}{2} \delta_
        +'\big]^\top, 
  $
    \label{eq:second_ders}
\end{small}
where we defined
\begin{small}
    $\delta'_{\pm}  := \sigma_{\pi}^{\prime\prime}(\nd{v}_{\text{com}} + \frac{\nd{v_s}}{2}) \pm \sigma_{\pi}^{\prime\prime}(\nd{v}_{\text{com}} - \frac{\nd{v_s}}{2}),$
\end{small}
\end{subequations}
and denoted $\frac{d^2 \sigma_\pi}{dx^2} = \sigma_\pi''(x) = -\frac{2 \pi_\epsilon^2}{1+\tanh(\nd{n_f})} \text{sech}^2(\pi_\epsilon x + \nd{n_f}) \tanh(\pi_\epsilon x + \nd{n_f}) = -2 \pi_\epsilon \sigma'_\pi(x) \tanh(\pi_\epsilon x + \nd{n_f})$. 

\subsubsection{Third order derivatives}
\label{subsubsec:third_order_ders}
Proceeding analogously,
\begin{subequations}
\begin{small}
$
    \frac{\partial^3 \boldsymbol{f}}{\partial \nd{V}^3}(\boldsymbol{\nd{x}})  = \big[ -6 \pi_\nd{c} \;\; 0\;\; 0 \;\; 0\big]^\top, \;
    \frac{\partial^3 \boldsymbol{f}}{\partial \nd{v_{com}}^3}(\boldsymbol{\nd{x}})  = \big[ 0\;\; -\frac{\pi_{\nd{f}}}{2} \delta_+'' \;\; 0 \;\; -\pi_{\nd{f}} \delta_-'' \big]^\top,  \;
          \frac{\partial^3 \boldsymbol{f}}{\partial \nd{v_s^3}}(\boldsymbol{\nd{x}})  = \big[ 0\;\; -\frac{\pi_{\nd{f}}}{16} \delta_-'' \;\; 0 \;\; -\frac{\pi_{\nd{f}}}{8} \delta_+'' \big]^\top, \;
     \frac{\partial^3 \boldsymbol{f}}{\partial \nd{v_s^2}\partial \nd{v_{com}}}(\boldsymbol{\nd{x}})   = \big[ 0\;\; -\frac{\pi_{\nd{f}}}{8} \delta_+'' \;\; 0 \;\; -\frac{\pi_{\nd{f}}}{4} \delta_-'' \big]^\top,  \;
       \frac{\partial^3 \boldsymbol{f}}{\partial \nd{v_s}\partial \nd{v_{com}}^2}(\boldsymbol{\nd{x}})  = \big[ 0\;\; -\frac{\pi_{\nd{f}}}{4} \delta_-'' \;\; 0 \;\; -\frac{\pi_{\nd{f}}}{2} \delta_+'' \big]^\top,  
$
\label{eq:third_order_ders}
\end{small} 
where we defined
\begin{small}
$
    \delta_{\pm}''  := \sigma_{\pi}^{\prime\prime\prime}(\nd{v}_{\text{com}} + \frac{\nd{v_s}}{2}) \pm \sigma_{\pi}^{\prime\prime\prime}(\nd{v}_{\text{com}} - \frac{\nd{v_s}}{2}),
$
\end{small}
\end{subequations}
with $\frac{d^3 \sigma_\pi}{d x^3} = \sigma_\pi'''(x)= -2 \pi_\epsilon \big( \sigma_\pi''(x) \tanh(\pi_\epsilon x + \nd{n_f}) + \pi_\epsilon \sigma_\pi'(x) \text{sech}^2(\pi_\epsilon x + \nd{n_f})\big).$

\begin{rem}
\label{rem:useful_identities_FPs}
At all fixed points of \eqref{eq:dimensionless_electromechanical_dynamics_movingFrame}: $\delta_+ = 2 \sigma_\pi'(0)$, $\delta_+' = 2 \sigma_\pi''(0)$, $\delta_+ '' = 2 \sigma_\pi'''(0)$ and $\delta_- = \delta_-' = \delta_-'' = 0$.
\end{rem}

\subsection{Proof of Proposition \ref{prop:bistability}}
\label{subsec:proof_bistability}
We show the local asymptotic stability of $\boldsymbol{\nd{x}}_\pm$ in the  regime $0< \pi_{\nd{s}}< \pi_\nd{s}^\nd{H}$. The eigenvalues $\lambda_i$ of $\nabla_{\boldsymbol{\nd{x}}} \boldsymbol{f}(\boldsymbol{\nd{x}}_\pm; \pi_\nd{s})$ are continuous functions of $\pi_\nd{s}$. This follows from \cite[Theorem 3.9.1]{Tyrtyshnikov1997}, by noticing that the coefficients of the characteristic polynomial  \eqref{eq:char_poly_Hopf} are continuous in $\pi_\nd{s}$.  Furthermore, at $\pi_\nd{s} = 0$, $\nabla_{\boldsymbol{\nd{x}}} \boldsymbol{f}(\boldsymbol{\nd{x}}_\pm; 0)$ has eigenvalues 
$
\lambda_1 = -2 \pi_{\nd{l}}, 
\lambda_2  = - \pi_{\nd{f}} \sigma_{\pi}'(0), 
\lambda_{3,4}  =  \frac{- \pi_{\nd{f}} \sigma_{\pi}'(0) - 2 \zeta \pm \sqrt{(\pi_{\nd{f}} \sigma_{\pi}'(0) + 2 \zeta)^2 - 4} }{2} . 
$ \eqref{eq:char_poly_Hopf} factorizes providing the eigenvalue $\lambda = - \pi_\nd{f} \sigma_\pi'(0)$, independent of $ \pi_\nd{s}$, which is strictly negative real under Assumption \ref{ass:parameter_ranges}. The roots of $p_3$ in \eqref{eq:char_poly_Hopf} provide $\pi_\nd{s}$-dependent eigenvalues. As $p_3$ is a cubic polynomial with real coefficients, either it has three real roots or a real root and two complex conjugate roots. Thus, the eigenvalues can lose stability as $\pi_\nd{s}$ increases from 0 only in two ways: (1) At least one real eigenvalue becomes zero, transitioning from the strictly negative real line to the strictly positive. 
        For this to happen, $p_3$ must have a factor $\lambda \, \Rightarrow \, \pi_\nd{s} \ = \frac{\pi_\nd{l}}{2 \pi_\nd{V}}$ (i.e., $\nd{c}_0$ in $p_3$ must vanish).
        However, $\pi_\nd{s}^{\nd{H}} < \frac{\pi_\nd{l}}{2 \pi_\nd{V}}$, and thus, a real eigenvalue cannot transition from the negative to the positive real line in the parameter regime $0<\pi_\nd{s} <\pi_\nd{s}^{\nd{H}}$. (2) Two conjugate complex eigenvalues transition from the left-half  to the right-half complex plane.
        This is the scenario analyzed in Theorem \ref{thm:equivariant_Hopf} and takes place at $\pi_\nd{s}  = \pi_\nd{s}^{\nd{H}}$.
    Consequently, the eigenvalues of $\nabla_{\boldsymbol{\nd{x}}} \boldsymbol{f}(\boldsymbol{\nd{x}}_\pm; \pi_\nd{s})$ remain in the open left-half complex plane for $0< \pi_\nd{s} <\pi_\nd{s}^\nd{H} \; \Rightarrow \; \boldsymbol{\nd{x}}_{\pm}$ are asymptotically stable in this parameter regime by Lyapunov's linearization method \cite[Theorem 3.1]{Slotine1991}.

\subsection{Proof of Theorem \ref{thm:equivariant_Hopf}: Hopf bifurcation at the symmetric fixed points $\boldsymbol{\nd{x}_\pm}$}
\label{subsubsec:proof_equivariant_Hopf}

We proceed with the proof of the Hopf bifurcation in the excitable crawler by applying Theorem \ref{thm:Hopf_bifurcation} to the dynamics \eqref{eq:dimensionless_electromechanical_dynamics_movingFrame} at the critical value $\pi_\nd{s}^\nd{H}$ as defined in \eqref{eq:parameter_hopf_1}. The characteristic polynomial of the Jacobian $\nd{J}^\nd{H}:= \nabla_{\boldsymbol{\nd{x}}} \boldsymbol{f}(\boldsymbol{\nd{x}}_{\pm}; \pi_{\nd{s}})$ is
\begin{equation}
    \big( \pi_{\nd{f}}  \, \sigma_{\pi}^{\prime}(0) + \lambda  \big)  \, \underbrace{\Big(\lambda^3 + \nd{c_2} \lambda^2 + \nd{c_1}  \lambda + \nd{c_0}\Big)}_{=: p_3(\lambda)} = 0, 
    \label{eq:char_poly_Hopf}
\end{equation}
with
 $\nd{c_2}(\pi_\nd{s})  := \big( \gamma - 2 (3 \pi_{\nd{s}} \pi_{\nd{V}} - \pi_{\nd{l}}) \big), \;
    \nd{c_1}(\pi_\nd{s})  := \big( 1 - 2 \gamma (3 \pi_{\nd{s}} \pi_{\nd{V}} - \pi_{\nd{l}}) \big), \text{ and }
    \nd{c_0}(\pi_\nd{s})  = 2 ( \pi_{\nd{l}} - 2 \pi_{\nd{s}} \pi_{\nd{V}})$. From \eqref{eq:char_poly_Hopf}, an eigenvalue of $\nd{J}^\nd{H}$ is $\lambda_1 = - \pi_{\nd{f}} \, \sigma'_{\pi}(0) <0$, independently of the value of $\pi_\nd{s}$. We show next that $\pi_\nd{s} = \pi_\nd{s}^\nd{H}$ is the value at which the cubic polynomial $p_3(\lambda)$ in \eqref{eq:char_poly_Hopf} has a conjugate pair of purely imaginary roots. When a cubic polynomial with real coefficients has two purely imaginary roots ($\pm \beta \mathbf{i}, \beta \in \mathbb{R}$) and a real root ($\alpha \in \mathbb{R}$), its coefficients are of the following form
            \begin{align}
                (\lambda - \alpha)(\lambda - \beta \mathbf{i})(\lambda + \beta \mathbf{i}) = 
                 \lambda^3 - \alpha \lambda^2 + \beta^2 \lambda - \alpha \beta^2. 
                 \label{eq:Hopf_coeff_structure}
            \end{align}
Matching coefficients in \eqref{eq:char_poly_Hopf} and \eqref{eq:Hopf_coeff_structure} provides the identities
            \begin{subequations}
            \begin{align}
                \mathcal{O}(\lambda^2): \;\;\; & - \alpha  = \nd{c}_2(\pi_\nd{s}^\nd{H}),
                \label{eq:alpha}\\
                \mathcal{O}(\lambda): \;\;\; & \beta^2    = \nd{c_1}(\pi_\nd{s}^\nd{H}) \;
                \Rightarrow \; 
                \beta = \sqrt{\nd{c_1}(\pi_\nd{s}^\nd{H})}, 
                \label{eq:beta}\\
                \mathcal{O}(1): \;\;\; &  - \alpha \beta^2    = \nd{c_0}(\pi_\nd{s}^\nd{H}).
                \label{eq:compatibility_condition}
            \end{align}
             Since $\beta$ as defined in \eqref{eq:beta} is such that $\beta \in \mathbb{R}_{++}$:
             \begin{equation}
                 \pi_\nd{s}^\nd{H} \; <  \; \frac{1}{3 \pi_\nd{V}} \bigg( \frac{1}{2 \gamma} + \pi_\nd{l}\bigg).
                 \label{eq:cond4}
             \end{equation}
              \end{subequations}
            \eqref{eq:cond4} together with the bounds in $\gamma$  $\Rightarrow$
            $\alpha <0$, with $\alpha$ as defined in \eqref{eq:alpha}. 
\eqref{eq:compatibility_condition}-\eqref{eq:cond4} together yield a \textit{compatibility condition} defining the value $\pi_{\nd{s}}^\nd{H}$. Substituting \eqref{eq:alpha}-\eqref{eq:beta} into \eqref{eq:compatibility_condition} and rearranging terms:
\begin{small}
            \begin{equation}
               \underbrace{ 1 - 2 \gamma \big(3 \pi_{\nd{s}}^\nd{H} \pi_{\nd{V}} - \pi_\nd{l} \big) }_{>0 \text{ by } \eqref{eq:cond4}} = \frac{2 (\pi_\nd{l} - 2 \pi_{\nd{s}}^\nd{H} \pi_{\nd{V}})}{\gamma - 2 (3 \pi_{\nd{s}}^\nd{H} \pi_{\nd{V}} - \pi_{\nd{l}})}.
               \label{eq:Hopf_cond2}
            \end{equation}
            \end{small}
            
            Define $\nd{f}(\pi_\nd{s}) := 1 - 2 \gamma \big(3 \pi_{\nd{s}} \pi_{\nd{V}} - \pi_\nd{l} \big)$ and $\nd{g}(\pi_\nd{s}) = \frac{2 (\pi_\nd{l} - 2 \pi_{\nd{s}} \pi_{\nd{V}})}{\gamma - 2 (3 \pi_{\nd{s}} \pi_{\nd{V}} - \pi_{\nd{l}})}$. Then, $\pi_\nd{s}^\nd{H}$ is determined by the intersection of $\nd{f}$ and $\nd{g}$ and must satisfy
            $
                0 < \pi_\nd{s}^\nd{H} < \frac{1}{3 \pi_\nd{V}} \big( \frac{1}{2 \gamma}  + \pi_\nd{l} \big).
                \label{eq:Omega_bounds}
           $
Further note that:
  $  \nd{f}'(\pi_\nd{s}) = - 6 \gamma \pi_\nd{V} < 0 \;\; \Rightarrow \; \nd{f}$ is monotonically decreasing and  $\nd{f}(0 ) = 1 + 2 \gamma \pi_\nd{l}>1$. $\nd{g}'(\pi_\nd{s})  = 4 \pi_\nd{V} \; \frac{(\pi_\nd{l} - \gamma)}{\big(\gamma^2 - 2 (3 \pi_\nd{s}\pi_\nd{V} - \pi_\nd{l})\big)^2} > 0 \; \Rightarrow \; \nd{g}$
is monotonically increasing and  $\nd{g}(0) = \frac{2 \pi_\nd{l}}{\gamma + 2 \pi_\nd{l}} < 1$. Since $\nd{f}$ is strictly decreasing and $\nd{g}$ is strictly increasing, 
there can be at most one intersection in the admissible interval. 
Moreover, $\nd{f}(0)>\nd{g}(0)$, whereas
$
\lim_{\pi_\nd{s} \to
\frac{1}{3\pi_\nd{V}}\left(\frac{1}{2\gamma}+\pi_\nd{l}\right)^-}
\nd{f}(\pi_\nd{s})=0,
$
while $\nd{g}(\pi_\nd{s})>0$ throughout this interval. Hence, by continuity,
$\nd{f}$ and $\nd{g}$ have exactly one intersection in the admissible interval.

 Define $\Omega_\nd{H}:= \pi_\nd{s}^\nd{H} \pi_\nd{V}$ and rearranging terms in \eqref{eq:Hopf_cond2},
\begin{small}
           $
               \Omega_\nd{H}^2 - \Big( \frac{1}{6} \gamma + \frac{1}{18 \gamma} + \frac{2}{3} \pi_{\nd{l}}\Big) \Omega_\nd{H} + \frac{1}{36} \Big( 4 \pi_{\nd{l}}^2 + 2 \gamma \pi_{\nd{l}} + 1  \Big) = 0,
        \label{eq:quadratic_for_Hopf_param}
           $
\end{small}
which has solutions
\begin{small}
             \begin{equation}
                \Omega_\nd{H}^{\pm} = \frac{1}{12} \gamma + \frac{1}{36 \gamma} + \frac{1}{3} \pi_{\nd{l}} \pm \frac{1}{2}\sqrt{\frac{1}{18^2 \gamma^2 } + \frac{1}{36 }\gamma^2 + \frac{2 \pi_\nd{l}}{27 \gamma} -\frac{5}{54} }.
                \label{eq:sols}
            \end{equation}
\end{small}    
The bounds assumed on $\gamma$ ensure the positivity of the radicand in \eqref{eq:sols} and $\Omega_\nd{H}^{\pm}>0$. Under such an assumption,  $\Omega_\nd{H}^{+}$---corresponding to  the solution with the $+$ sign in \eqref{eq:sols}---does \textit{not} satisfy the compatibility condition \eqref{eq:cond4}. Thus, the solution of interest is $\Omega_\nd{H}^{-}$ and the critical value of the bifurcation parameter is $\pi_\nd{s}^\nd{H} = \Omega_\nd{H}^{-}/\pi_\nd{V}$, as given in \eqref{eq:parameter_hopf_1}. We have shown that  $\nabla_{\boldsymbol{\nd{x}}} \boldsymbol{f}(\boldsymbol{\nd{x}}_\pm; \pi_\nd{s}^\nd{H})$ has two conjugate purely imaginary eigenvalues and the other two eigenvalues are real and negative. Next, we compute the speed of the conjugate complex pair of eigenvalues of interest as they cross the imaginary axis. Defining $\Omega := \pi_\nd{s} \pi_\nd{V}$ for convenience and implicitly differentiating $p_3$:
\begin{small}
\begin{align}
                \frac{\partial \lambda}{\partial \Omega} = \frac{6 \lambda^2 + 6 \gamma \lambda + 4 }{3 \lambda^2 + 2 \lambda \big( \gamma - 2 (3 \Omega - \pi_{\nd{l}})  \big) + 1 - 2 \gamma(3\Omega - \pi_{\nd{l}}) }.
                \label{eq:lambda_der}
            \end{align}
            \end{small}
Evaluating \eqref{eq:lambda_der} at $\Omega_\nd{H}$ and the eigenvalues of interest (i.e.,  $\pm \beta \mathbf{i}$) and after some algebraic manipulations 
            $
                \frac{\partial \lambda}{\partial \Omega} \big\vert_{\Omega_{\nd{H}}}
                = \frac{-3 \beta^2 + 3 \gamma \beta \mathbf{i} + 2}{\beta \Big(- \beta +   \mathbf{i} \big(   \gamma - \frac{(1-\beta^2)}{\gamma}\big)\Big) },
                \label{eq:eig_der_evaluation}
          $
          where $\beta$ is as given in \eqref{eq:beta}. Since only the rate of the \textit{real} part of the eigenvalues is of interest, 
          \begin{small}
          $
              \frac{\partial \Re(\lambda)}{\partial \Omega} \big\vert_{\Omega_{\nd{H}}}= \frac{ 3 \gamma^2 +6 \beta^2- 5}{\beta^2  + \big(   \gamma - \frac{(1-\beta^2)}{\gamma}\big)^2 } >0
              \label{eq:real_rate}
        $
          \end{small}
---positivity follows from the fact that
$\beta^2= \nd{f}(\pi_\nd{s}^\nd{H})=\nd{g}(\pi_\nd{s}^\nd{H})$, and since $\nd{g}$ is strictly increasing
and $\pi_\nd{s}^\nd{H}>0$,
$\beta^2>\nd{g}(0)=\frac{2\pi_\nd{l}}{\gamma+2\pi_\nd{l}}
>\frac{2}{3},
$
where the last inequality follows from $\pi_\nd{l}>\gamma$. Therefore,
$
3\gamma^2+6\beta^2-5
>
3\gamma^2-1>0,
$
since $\gamma>1$. Since $\pi_\nd{V}>0$, $\mathrm{sign}\big(\partial_\Omega \Re(\lambda)\big) = \text{sign}\big(\partial_{\pi_{\nd{s}}} \Re(\lambda)\big)$ and the complex conjugate pair of eigenvalues transition from the left-half to the right-half complex plane as $\pi_\nd{s}$ is increased through $\pi_\nd{s}^\nd{H}$. This completes the linear part of the analysis. To determine the criticality of the bifurcation, we analyze the coefficient of the cubic term of the Hopf normal form---the first Lyapunov coefficient, $l_1$. Following \cite[\S 5.4.1]{Kuznetsov2023} the eigenvectors of interest are
       \begin{small}
       $
           \boldsymbol{q} = 
           \begin{bmatrix}
               \frac{\pi_\nd{s}^\nd{H}}{2(3 \Omega_\nd{H} - \pi_\nd{l}) - \beta \mathbf{i}} \;\;
               0\;\;
               1\;\;
               \beta \mathbf{i}
           \end{bmatrix}^\top  q_3 \text{ and }
           \boldsymbol{p} = 
           \begin{bmatrix}
                  \frac{-2\pi_\nd{V}}{2(3 \Omega_\nd{H} - \pi_\nd{l}) + \beta \mathbf{i}}\;\;
               0\;\;
               \gamma - \mathbf{i} \beta\;\;
               1
           \end{bmatrix}^\top p_4
       $
       \end{small}
        with 
\begin{small}
$
\Big( \frac{-2 \Omega_\nd{H}}{\big(2(3 \Omega_\nd{H} - \pi_\nd{l}) - \beta \mathbf{i}\big)^2} + \gamma + 2 \beta \mathbf{i} \Big) q_3 \bar{p}_4
= 1.$
\end{small}
For notational compactness, we define $\kappa:=4(3 \Omega_\nd{H} - \pi_\nd{l})^2 + \beta^2>0$. Leveraging sparsity in the derivatives in \hyperlink{sec:Appendix_II}{Appendix I.A} evaluated at $(\boldsymbol{\nd{x}}_\pm; \pi_\nd{s}^\nd{H})$ and using symbolic computation tools, we obtain the inner products in the definition of the first Lyapunov coefficient  
\begin{small}
$
     l_{1a}:= -2 \langle \boldsymbol{p}, \mathrm{d}^2\boldsymbol{f}_{(\boldsymbol{\nd{x}}_\pm; \pi_\nd{s}^\nd{H})}\Big[\boldsymbol{q}, \nabla_{\boldsymbol{\nd{x}}} \boldsymbol{f}(\boldsymbol{\nd{x}}_\pm; \pi_\nd{s}^\nd{H})^{-1} \mathrm{d}^2\boldsymbol{f}_{(\boldsymbol{\nd{x}}_\pm; \pi_\nd{s}^\nd{H})}[\boldsymbol{q}, \bar{\boldsymbol{q}}]\Big] \rangle  = 
      q_3 |q_3|^2 \bar{p}_4 \bigg( \frac{-72 \pi_\nd{V} \pi_\nd{c} (\pi_\nd{s}^{\nd{H}})^3}{[2(3 \pi_\nd{V} \pi_\nd{s}^\nd{H}-\pi_\nd{l}) - \beta \mathbf{i}]^2 \kappa}+ \frac{\pi_\nd{f}}{2} \frac{\big( \sigma_\pi''(0)\big)^2}{\sigma_\pi'(0)} \beta^3 \mathbf{i}\bigg); 
    \;\; l_{1b}:=  \langle \boldsymbol{p}, \mathrm{d}^3 \boldsymbol{f}_{(\boldsymbol{\nd{x}}_\pm; \pi_\nd{s}^\nd{H})}[\boldsymbol{q},\boldsymbol{q}, \bar{\boldsymbol{q}}] \rangle = 
      q_3 |q_3|^2 \bar{p}_4  \bigg( \frac{12 \pi_\nd{V} \pi_\nd{c} (\pi_\nd{s}^\nd{H})^3}{\kappa \big( 2(3 \pi_\nd{V} \pi_\nd{s}^\nd{H}-\pi_\nd{l}) - \beta \mathbf{i}\big)^2} - \frac{\pi_\nd{f}}{4} \sigma'''_\pi(0) \beta^3 \mathbf{i}\bigg);$ 
      \end{small}
      and 
      \begin{small}
      $
     l_{1c}:= \langle \boldsymbol{p}, \mathrm{d}^2\boldsymbol{f}_{(\boldsymbol{\nd{x}}_\pm; \pi_\nd{s}^\nd{H})}\Big[\bar{\boldsymbol{q}}, \big( 2\beta \mathbf{i} \mathbb{I}_4 - \nabla_{\boldsymbol{\nd{x}}} \boldsymbol{f}(\boldsymbol{\nd{x}}_\pm; \pi_\nd{s}^\nd{H})\big)^{-1} \mathrm{d}^2\boldsymbol{f}_{(\boldsymbol{\nd{x}}_\pm; \pi_\nd{s}^\nd{H})}[\boldsymbol{q}, \boldsymbol{q}]\Big] \rangle = -6 \pi_\nd{c} \sqrt{\frac{\pi_\nd{l} - 2 \pi_\nd{V}\pi_\nd{s}^\nd{H}}{\pi_\nd{c}}} \bar{q}_1 \bar{p}_1 h_1 - \pi_\nd{f} \sigma_\nd{\pi}''(0) \bar{q}_4 \bar{p}_4 h_2,  
$
\end{small}
where $p_i$ (resp. $q_i$) denotes the $i$-th entry of the vector $\boldsymbol{p}$ (resp. $\boldsymbol{q}$) and $
h_2 = -\frac{\pi_\nd{f}}{4}\, \frac{\sigma_\pi''(0)}{2\mathbf{i}\beta+\pi_\nd{f}\sigma_\pi'(0)}\, q_4^{\,2}$, $
h_1 = \frac{1-4\beta^2+2\mathbf{i}\beta\gamma}{2\pi_\nd{V}}\, h_3$ with $
h_3 = \frac{-6\pi_\nd{c} \nd{V}_\pm q_1^2}{\pi_\nd{s}^\nd{H}
+ \frac{(2\mathbf{i}\beta-\pi_\nd{l}-3\pi_\nd{c} \nd{V}_\pm^2)(1-4\beta^2+2\mathbf{i}\beta\gamma)}{2\pi_\nd{V}}
}$.
The first Lyapunov coefficient is
\begin{equation}
    l_1 := \Re\big(l_{1a} + l_{1b} +l_{1c} \big).
    \label{eq:l1}
\end{equation}
Whenever $l_1 \neq 0$, the Hopf bifurcation is non-degenerate. When $l_1 < 0$ (resp., $l_1 >0$), the Hopf bifurcation is supercritical (resp., subcritical).

\vspace{-0.5cm}
\subsection{Proof of Theorem \ref{thm:pitchfork_crawler}: pitchfork bifurcation at $\boldsymbol{\nd{x}}_0$}
\label{subsubsec:pitchfork_proof}

The proof follows by an application of Theorem \ref{thm:pitchfork_bifurcation} to the excitable crawler dynamics \eqref{eq:dimensionless_electromechanical_dynamics_movingFrame} at the critical parameter value $\pi_\nd{s}^\nd{P}  = \frac{\pi_\nd{l}}{2 \, \pi_\nd{V}}$. The characteristic polynomial of   $\nabla_{\boldsymbol{\nd{x}}} \boldsymbol{f}(\boldsymbol{\nd{x}}_0; \pi_\nd{s}^\nd{P}  = \frac{\pi_\nd{l}}{2 \pi_\nd{V}}) =: J$, 
is
    $- \big(\pi_\nd{f} \sigma'_\pi(0)+ \lambda \big) \lambda  \big( (\pi_\nd{l} - \lambda) (\gamma + \lambda) - 1 \big)  = 0. $ The corresponding eigenvalues are
 $\lambda_1  = - \pi_\nd{f} \sigma'_\pi(0) < 0,
        \lambda_2 = 0, \text{ and }
        \lambda_{3,4}  = \frac{\pi_\nd{l} - \gamma \pm \sqrt{(\gamma + \pi_\nd{l})^2 - 4}}{2} \neq 0$ since $\gamma \pi_\nd{l} > 1$ by assumption.
Thus, there is a unique zero eigenvalue. Left and right eigenvectors for $\lambda_2 = 0$ are
    $J \boldsymbol{v}  = \boldsymbol{0}\; \Rightarrow \; \boldsymbol{v} = \big[ 1 \;\; 0 \;\; 2 \pi_\nd{V} \;\; 0\big]^\top \label{eq:pitchfork_right_vector}$
    and
    $\boldsymbol{w}^\top J  = \boldsymbol{0}\; \Rightarrow \; \boldsymbol{w} = \frac{1}{\gamma \pi_\nd{l}-1}\big[ - 1 \;\; 0 \;\; \gamma \pi_\nd{s}^\nd{P} \;\; \pi_\nd{s}^\nd{P} \big]^\top \label{eq:pitchfork+left_vector}$, respectively---we normalized $\boldsymbol{w}$ such that $\langle \boldsymbol{w}, \boldsymbol{v} \rangle = 1$. We have checked conditions \textit{(1)-(3).} \textit{(4)} Differentiation with respect to the bifurcation parameter yields
$
    \partial_{\pi_{\nd{s}}} \boldsymbol{f}(\boldsymbol{\nd{x}})  = \big[ -\nd{s} \;\; 0 \;\; 0 \;\; 0 \big]^\top \;\; \Rightarrow \;\;\partial_{\pi_{\nd{s}}} \boldsymbol{f}(\boldsymbol{\nd{x}}_0) = \boldsymbol{0}. 
$
Consequently, $\langle \boldsymbol{w}, \partial_{\pi_{\nd{s}}} \boldsymbol{f}(\boldsymbol{\nd{x}}_0) \rangle = 0$.   \textit{(5)} The only non-zero element of the tensor $\partial_{\boldsymbol{\nd{x}}} \big( \partial_{\pi_{\nd{s}}} \boldsymbol{f}(\boldsymbol{\nd{x}})\big)$ is $\partial_{\nd{s}} \big( \partial_{\pi_{\nd{s}}} f_1(\boldsymbol{\nd{x}})\big) = -1.$ Accordingly, $\mathrm{d}  ( \partial_{\pi_{\nd{s}}} \boldsymbol{f})_{(\boldsymbol{\nd{x}}_0; \pi_\nd{s}^\nd{P}})[\boldsymbol{v}] = [-2 \pi_\nd{V} \;\; 0 \;\; 0 \;\; 0]^\top$ and $\langle \boldsymbol{w}, \mathrm{d}  ( \partial_{\pi_{\nd{s}}} \boldsymbol{f})_{(\boldsymbol{\nd{x}}_0; \pi_\nd{s}^\nd{P}})[\boldsymbol{v}] \rangle = \frac{2 \pi_\nd{V}}{\gamma \pi_\nd{l}-1} \neq 0$ (since $\partial_{\pi_{\mathsf{s}}}\boldsymbol{f}(\boldsymbol{\nd{x}}_0;\pi_{\mathsf{s}}^\nd{P})=\mathbf{0}$, the correction term in condition \textit{(5)} involving $J^{-1}P(\partial_{\pi_{\mathsf{s}}}\boldsymbol{f})$ vanishes). \textit{(6)} Evaluating second-order derivatives at $\boldsymbol{\nd{x}}_0$ yields:
\begin{small}
$
     \frac{\partial^2 \boldsymbol{f}}{\partial \nd{V}^2}(\boldsymbol{\nd{x}}_0)  = \boldsymbol{0}, 
    \frac{\partial^2 \boldsymbol{f}}{\partial \nd{v_{com}}^2}(\boldsymbol{\nd{x}}_0)   = \big[ 0\;\; -\pi_{\nd{f}}\sigma_\pi''(0)  \;\; 0 \;\; 0  \big]^\top, 
     \frac{\partial^2 \boldsymbol{f}}{\partial \nd{v_{s}}^2}(\boldsymbol{\nd{x}}_0)  = \big[ 0\;\; -\frac{\pi_{\nd{f}}}{4}\sigma_\pi''(0)  \;\; 0 \;\; 0  \big]^\top,   
     \frac{\partial^2 \boldsymbol{f}}{\partial \nd{v_{com}} \partial \nd{v_s} }(\boldsymbol{\nd{x}}_0)  = \big[ 0\;\; 0  \;\; 0 \;\; -\pi_{\nd{f}}\sigma_\pi''(0) \big]^\top = 
     \frac{\partial^2 \boldsymbol{f}}{\partial \nd{v_s} \partial \nd{v_{com}}}(\boldsymbol{\nd{x}}_0).
\label{eq:pitchform_second_ders}
$
\end{small}
Thus, 
\begin{small}
$    \mathrm{d}^2 \boldsymbol{f}_{
             (\boldsymbol{\nd{x}}_0, \pi_\nd{s}^\nd{P})}[\boldsymbol{v},\boldsymbol{v}]  
             = \boldsymbol{0} 
$
\end{small}
$\Rightarrow \langle \boldsymbol{w}, \mathrm{d}^2 \boldsymbol{f}_{
             (\boldsymbol{\nd{x}}_0, \pi_\nd{s}^\nd{P})}[\boldsymbol{v},\boldsymbol{v}]\rangle = 0.$  \textit{(7)} Finally, third-order derivatives at $\boldsymbol{\nd{x}}_0$:
\begin{small}
$
    \frac{\partial^3 \boldsymbol{f}}{\partial \nd{V}^3}(\boldsymbol{\nd{x}}_0)  = \big[ -6 \pi_\nd{c} \;\; 0\;\; 0 \;\;  0\big]^\top,
     \frac{\partial^3 \boldsymbol{f}}{\partial \nd{v_{com}}^3}(\boldsymbol{\nd{x}}_0)  = \big[ 0 \;\; - \pi_\nd{f} \sigma'''_\pi(0)\;\; 0 \;\; 0\big]^\top,  
     \frac{\partial^3 \boldsymbol{f}}{\partial \nd{v_s}\partial \nd{v_{com}}^2}(\boldsymbol{\nd{x}}_0)  = \big[ 0 \;\; 0\;\; 0 \;\; - \pi_\nd{f} \sigma'''_\pi(0)\big]^\top, 
    \frac{\partial^3 \boldsymbol{f}}{\partial \nd{v_s}^2 \partial \nd{v_{com}}}(\boldsymbol{\nd{x}}_0)  = \big[ 0 \;\; - \frac{\pi_\nd{f}}{4} \sigma'''_\pi(0) \;\; 0 \;\; 0\big]^\top,  
   \frac{\partial^3 \boldsymbol{f}}{\partial \nd{v_s}^3 }(\boldsymbol{\nd{x}}_0)  = \big[ 0 \;\; 0 \;\; 0 \;\; -\frac{\pi_\nd{f}}{4} \sigma'''_\pi(0) \big]^\top. 
$
\end{small}
The third-order directional derivative of interest is given by 
\begin{small}
$    \mathrm{d}^3 \boldsymbol{f}_{
               (\boldsymbol{\nd{x}}_0, \pi_\nd{s}^\nd{P})}[\boldsymbol{v},\boldsymbol{v},\boldsymbol{v}] 
               = \big[ -6 \pi_\nd{c} \;\; 0 \;\; 0 \;\;  0\big]^\top.
$
\end{small} Projecting onto $\boldsymbol{w}$, 
\begin{small}
 $  \langle \boldsymbol{w},   \mathrm{d}^3 \boldsymbol{f}_{
               (\boldsymbol{\nd{x}}_0, \pi_\nd{s}^\nd{P})}[\boldsymbol{v},\boldsymbol{v},\boldsymbol{v}] \rangle  
                = \frac{6 \pi_\nd{c} }{ \gamma \pi_\nd{l}-1} \neq  0 
$
\end{small}
(the remaining inner product in \textit{(7)} vanishes.) Since all the conditions in Theorem \ref{thm:pitchfork_bifurcation} are met, the excitable crawler undergoes a \textit{pitchfork bifurcation} at $(\boldsymbol{\nd{x}}_0, \pi_\nd{s}^\nd{P})$. 

\textit{Criticality \& stability.} After reduction to the center manifold, the dynamics are of the form $u' = \alpha_1 (\pi_\nd{s} - \pi_\nd{s}^\nd{P})u + \alpha_2 u^3 + h.o.t.$, where $\alpha_1$ has the same sign of \textit{(5)} and $\alpha_2$ the same sign of \textit{(7)}. Thus, non-trivial equilibria of the reduced dynamics are $u_\pm \approx \sqrt{\frac{-\alpha_1(\pi_\nd{s}-\pi_\nd{s}^\nd{P})}{\alpha_2}}$. Since $\alpha_1/\alpha_2 >0$ by assumption \ref{ass:parameter_ranges}, $u_{\pm}$ exist for $\pi_\nd{s}<\pi_\nd{s}^\nd{P}$. By assumption $\gamma \pi_\nd{l} > 1 \, \Rightarrow \, \alpha_1>0 \, \Rightarrow \, \partial_u u'\vert_{u_{\pm}}>0$, and thus, the bifurcating branches are 
locally unstable near $\pi_\nd{s}^\nd{P}$ on the one-dimensional center manifold. \hfill $\blacksquare$

\section*{Appendix II: Proof of Proposition \ref{prop:canards}}
\label{sec:Appendix_II}

\textit{a)} The \textit{folds} of the critical manifold $\mathcal{S}_0$ are the subsets of the manifold where the fast dynamics lose normal hyperbolicity \cite[Definition 3.4]{Wechselberger2013}. In the excitable crawler dynamics, they are characterized by \eqref{eq:folds}, corresponding to the local minima and maxima of the cubic critical manifold. 

\textit{b)} Implicit differentiation of the critical manifold yields $\nd{V}^\prime = \frac{\pi_\nd{s}^{(\varepsilon)} \nd{v}_\nd{s}}{-3 \pi_\nd{c}^{(\varepsilon)} \nd{V}^2 + \pi_\nd{l}^{(\varepsilon)}}$.   Furthermore, in the critical manifold, $\nd{s} =   \big( - \pi_{\nd{c}}^{(\varepsilon)} \,  \nd{V}^3 +\pi_{\nd{l}}^{(\varepsilon)}\, \nd{V} \big)/ \pi_{\nd{s}}^{(\varepsilon)}$.   Projecting the dynamics \eqref{eq:slowDynamics_singularPerturbation} 
into the coordinate chart $(\nd{V}, \nd{v}_\nd{com}, \nd{v}_\nd{s})\in \mathbb{R}^3$ yields an alternative representation of the reduced  dynamics that
blows up close to the folds \eqref{eq:folds}. 
We \textit{desingularize} the problem \cite{Wechselberger2013} by introducing the state-dependent time re-scaling $d\nd{t} =: - (-3 \pi_\nd{c}^{(\varepsilon)} \nd{V}^2 + \pi_\nd{l}^{(\varepsilon)} ) d \tau$.  
The corresponding \textit{desingularized slow dynamics} are
\begin{small}
\begin{subequations}
\begin{align}
     \frac{d \nd{V}}{d \tau} & = -\pi_\nd{s}^{(\varepsilon)} \nd{v}_\nd{s},\\
     \frac{d \nd{v}_{\text{com}}}{d \tau} &=   \frac{\pi_{\nd{f}}}{2} (\pi_\nd{l}^{(\varepsilon)} -3 \pi_\nd{c}^{(\varepsilon)} \nd{V}^2 )  \big( \sigma_{\pi}^+ + \sigma_{\pi}^-  \big), \label{eq:V_critical_des}\\
   \frac{d \nd{v_s}}{d \tau} &=  (3 \pi_\nd{c}^{(\varepsilon)} \nd{V}^2 - \pi_\nd{l}^{(\varepsilon)}) \Big(\pi_{\nd{f}} \big( \sigma_{\pi}^- - \sigma_{\pi}^+ \big)  + \frac{  \pi_{\nd{c}}^{(\varepsilon)} \,  \nd{V}^3 - \pi_{\nd{l}}^{(\varepsilon)} \, \nd{V} }{ \pi_{\nd{s}}^{(\varepsilon)}} \nonumber \\
   & \hspace{4.4cm}   - 2 \, \zeta \, \nd{v_s}   + 2 \pi_{\nd{v}} \nd{V}\Big).
\end{align}
\label{eq:desingularized_dynamics}
\end{subequations}
\end{small}
Due to the time re-scaling, the reduced flow is obtained from the desingularized flow by changing
the direction of the flow on the repelling branch. 
Equilibrium points of \eqref{eq:desingularized_dynamics} positioned at the folds of the critical manifold are called \textit{folded singularities} \cite[Def. 3.7]{Wechselberger2013}. 
In our system, the generic folded singularities satisfy \eqref{eq:folded_singularities}.

\textit{c)} The classification of generic folded singularities is based on the two non-zero eigenvalues of the Jacobian of \eqref{eq:desingularized_dynamics} evaluated at the folded singularities.
This Jacobian has eigenvalues $\lambda_1 = 0$ and 
$
    \lambda_{2,3} = \pm 2\sqrt{  \pi_\nd{l}^{(\varepsilon)} \big( \frac{\pi_\nd{l}^{(\varepsilon)}}{3} - \pi_\nd{V} \pi_\nd{s}^{(\varepsilon)}\big)}.
$ 
When the two non-zero eigenvalues are real and of opposite sign, the folded singularities are \textit{folded saddles} \cite[Def. 3.9]{Wechselberger2013}. 
Consequently, in our system the folded singularities are \textit{folded saddles} when the parametric condition in Proposition \ref{prop:canards}c) is satisfied\footnote{The value $\nd{v}_\nd{com} = 0$ is not excluded from the set of folded singularities $M_\mathcal{F}$ since to obtain an ordinary singularity of \eqref{eq:slowDynamics_singularPerturbation} at the folded singularities the additional condition $\pi_\nd{s} = \pi_\nd{l}/3 \pi_\nd{V}$ would need to be satisfied, which is excluded by Proposition \ref{prop:canards}c, avoiding degenerate cases.}. \\
\indent \textit{d)}
We proceed by finding asymptotic scalings as $\varepsilon \to 0$. From \eqref{eq:FPs_sym}, $\nd{V}_{\pm}(\pi_\nd{s}^\nd{H}) = \pm \sqrt{\frac{\pi_\nd{l}-2 \pi_\nd{V} \pi_\nd{s}^\nd{H}}{\pi_\nd{c}}}$ which we aim to compare to the voltage at the fold $\nd{V}_{\pm}^\nd{F} = \pm \sqrt{\frac{\pi_\nd{l}}{3 \pi_\nd{c}}}$ as $\varepsilon \to 0$. From \eqref{eq:parameter_hopf_1}, $\pi_\nd{s}^\nd{H} = \frac{1}{\pi_\nd{V}}\big( \frac{\pi_\nd{l}}{3} + C_0 - \frac{1}{2} \sqrt{A_0 + B_0 \pi_\nd{l}}\big)$, where $C_0 := \frac{\gamma}{12} + \frac{1}{36 \gamma}  , A_0:= \frac{1}{18^2 \gamma^2} + \frac{\gamma^2}{36} - \frac{5}{54} \text{ and } B_0:=\frac{2}{27 \gamma}$ are constants not dependent on $\varepsilon$. Taking into account that $\pi_\nd{s}, \pi_\nd{l}$ and $\pi_\nd{c}$ are $\mathcal{O}(\varepsilon^{-1})$, $\frac{\pi_\nd{l}-2 \pi_\nd{V} \pi_\nd{s}^\nd{H}}{\pi_\nd{c}} \approx \frac{\pi_\nd{l}^{(\varepsilon)}}{3 \pi_\nd{c}^{(\varepsilon)}} +  \mathcal{O}(\varepsilon^{1/2}) \; \Rightarrow \; \nd{V}_{\pm}(\pi_\nd{s}^\nd{H}) = \nd{V}_{\pm}^\nd{F} + \mathcal{O}(\varepsilon^{1/2}) \Rightarrow  \nd{s}_{\pm}(\pi_\nd{s}^\nd{H}) = \nd{s}_{\pm}^{\nd{F}} + \mathcal{O}(\varepsilon^{1/2}) $. The scaling of the imaginary eigenvalues $\lambda_{2,3} = \pm \beta(\varepsilon) \mathbf{i}$ of the Jacobian at the Hopf bifurcation point is derived from \eqref{eq:beta} in 
\hyperref[subsubsec:proof_equivariant_Hopf]{Appendix I.C}, substituting $\pi_\nd{s}^{\nd{H}, (\varepsilon)} = \frac{\pi_\nd{l}^{(\varepsilon)}}{3 \pi_\nd{V}} + \mathcal{O}(\varepsilon^{1/2})$. After a Taylor expansion, $\beta(\varepsilon) = 
\Theta(\varepsilon^{-1/4})$. \hfill $\blacksquare$

\section*{Appendix III: Describing function analysis \& Harmonic Balance}
\label{sec:Appendix_III}

\subsection*{A. Fundamental harmonic of the voltage}
\label{subsec:fundamental_H_voltage}
The DC component of the voltage is 
$
\begin{small}
     \nd{V}_0  := \frac{1}{2 \pi} \int_0^{2\pi} \nd{V}(\omega \nd{t}) \mathrm{d}(\omega \nd{t}) 
         = \frac{1}{2\pi} \Big( \int_0^\theta \nd{M} \mathrm{d}(\omega \nd{t}) + \int_{\theta}^{\pi+\theta} -\nd{M}  \mathrm{d}(\omega \nd{t}) + \int_{\pi+\theta}^{2\pi} \nd{M} \mathrm{d}(\omega \nd{t}) \Big) = 0,
\end{small}
$
 where $\theta$ is as shown in Fig.  \hyperlink{fig:harmonic_balance_CL}{5\textbf{B}}.

The coefficient of the sinusoidal component is
$\begin{small}
         \nd{V}_{\nd{sin}} := \frac{1}{\pi} \int_0^{2\pi} \nd{V}(\omega \nd{t}) \sin(\omega \nd{t}) \mathrm{d}(\omega \nd{t}) 
        = \frac{1}{\pi} \Big( \int_0^\theta \nd{M} \sin(\omega \nd{t}) \mathrm{d}(\omega \nd{t}) + \int_{\theta}^{\pi+\theta} -\nd{M} \sin(\omega \nd{t})  \mathrm{d}(\omega \nd{t}) + \int_{\pi+\theta}^{2\pi} \nd{M} \sin(\omega \nd{t}) \mathrm{d}(\omega \nd{t}) \Big) = -\frac{4}{\pi} \nd{M} \cos(\theta) 
         = -\frac{4}{\pi} \nd{M} \sqrt{1 - \big(\frac{\beta}{\nd{S}}\big)^2},
\end{small}$ 
where $\theta$ is as represented\footnote{Since by Assumption \ref{assumption:s_beta} $\nd{S} \geq \beta \; \Rightarrow \; \theta \in [0, \frac{\pi}{2}]$.} in Fig. \hyperlink{fig:harmonic_balance_CL}{5\textbf{B}} and defined implicitly by $\nd{S} \sin(\theta) = \beta \; \Rightarrow \; \sin \theta = \beta/\nd{S}$ and thus, $\cos(\theta) = \sqrt{1 - (\beta/\nd{S})^2}$.
Similarly,
$
\begin{small}
     \nd{V}_{\nd{cos}}  := \frac{1}{\pi} \int_0^{2\pi} \nd{V}(\omega \nd{t}) \cos(\omega \nd{t}) \mathrm{d}(\omega \nd{t})  = \frac{1}{\pi} \Big( \int_0^\theta \nd{M} \cos(\omega \nd{t}) \mathrm{d}(\omega \nd{t}) + \int_{\theta}^{\pi+\theta} -\nd{M} \cos(\omega \nd{t})  \mathrm{d}(\omega \nd{t}) + \int_{\pi+\theta}^{2\pi} \nd{M} \cos(\omega \nd{t}) \mathrm{d}(\omega \nd{t}) \Big) 
         =  \frac{4}{\pi} \nd{M} \sin(\theta) \nonumber  =  \frac{4}{\pi} \nd{M} \frac{\beta}{\nd{S}}.
\end{small}
$

\subsection*{B. Harmonic balance in the excitable crawler}
\label{subsec:harmonic_balance}
Relating the strain rate and local speeds provides the first set of harmonic balance equations. By definition $\nd{s}'(\nd{t}) = \nd{u}_2'(\nd{t}) - \nd{u}_1'(\nd{t})$. Substituting \eqref{eq:strain_harmonics} and \eqref{eq:speed1_harmonic_approx} yields
\begin{small}
$
     \omega \nd{S} \cos(\omega \nd{t})  = \bar{\nd{v}}_2 +\tilde{\nd{v}}_2 \cos(\omega \nd{t} + \phi_2)  - \bar{\nd{v}}_1 - \tilde{\nd{v}}_1 \cos(\omega \nd{t} + \phi_1)
     = \bar{\nd{v}}_2 - \bar{\nd{v}}_1  +\tilde{\nd{v}}_2 \cos(\phi_2) \cos(\omega \nd{t})  -  \tilde{\nd{v}}_2 \sin(\phi_2) \sin(\omega \nd{t})  - \tilde{\nd{v}}_1 \cos(\phi_1) \cos(\omega \nd{t}) + \tilde{\nd{v}}_1 \sin(\phi_1) \sin(\omega \nd{t}). 
 \nonumber 
$
\end{small}
Harmonic balance yields 
\begin{small}
\begin{subequations}
    \begin{align}
        \bar{\nd{v}}_2 &=  \bar{\nd{v}}_1  & 
        \label{eq:balance1}\\
        \nd{S}\omega &= \tilde{\nd{v}}_2 \cos(\phi_2)- \tilde{\nd{v}}_1 \cos(\phi_1)   & 
        \label{eq:balance2}\\
        \tilde{\nd{v}}_1 \sin(\phi_1) &= \tilde{\nd{v}}_2 \sin(\phi_2)   & 
        \label{eq:balance3}
    \end{align}
\label{eq:balance_strain_rate}
\end{subequations}
\end{small}
In view of \eqref{eq:balance1}, from this point onward we denote $\bar{\nd{v}}:= \bar{\nd{v}}_1 =  \bar{\nd{v}}_2$. The strain dynamics \eqref{eq:s_dynamics}-\eqref{eq:vs_dynamics}  provide the next set of balance equations. Substituting \eqref{eq:strain_harmonics}, \eqref{eq:speed1_harmonic_approx}, \eqref{eq:voltage_fundamental_component} and using the fundamental harmonics as approximations for the frictional force and voltage yields
\begin{small}
$
   -\omega^2 \nd{S} \sin(\omega \nd{t})  = \frac{\pi_\nd{f}}{\pi} \Big( \pi \Delta - \arccos(\nd{a}_2)(1 + \Delta) +  2 (1+\Delta) \cos(\phi_2) \sqrt{1-\nd{a}_2^2}\cos(\omega \nd{t})  -2 (1+\Delta) \sin(\phi_2) \sqrt{1-\nd{a}_2^2} \sin(\omega \nd{t}) - \pi \Delta  + \arccos(\nd{a}_1) (1 + \Delta)  - 2 (1+\Delta) \cos(\phi_1) \sqrt{1-\nd{a}_1^2} \cos(\omega \nd{t}) + 2 (1+\Delta) \sin (\phi_1) \sqrt{1-\nd{a}_1^2} \sin(\omega \nd{t}) \Big) - \nd{S}\sin(\omega \nd{t}) - 2 \zeta \nd{S} \omega \cos(\omega \nd{t}) +   \frac{ 8 \pi_\nd{V} \nd{M}}{\pi} \Big( -\sqrt{1 - \Big(\frac{\beta}{\nd{S}}\Big)^2} \sin(\omega \nd{t}) + \frac{\beta}{\nd{S}} \cos(\omega \nd{t})\Big), 
$
\end{small}
where $\nd{a}_1 := \frac{\bar{\nd{v}}_1}{\tilde{\nd{v}}_1}$ and $\nd{a}_2 := \frac{\bar{\nd{v}}_2}{\tilde{\nd{v}}_2}$.
Balancing DC terms provides
$        \arccos(\nd{a}_1) =  \arccos(\nd{a}_2)
$
$\Rightarrow \; \nd{a}_1 = \nd{a}_2 =: \nd{a}$. Together with \eqref{eq:balance1} $\Rightarrow \; \tilde{\nd{v}}_1 = \tilde{\nd{v}}_2 =: \tilde{\nd{v}}$. Substitution in \eqref{eq:balance3} yields $\sin(\phi_1) = \sin(\phi_2)$. Thus, either $\phi_1 = \phi_2$ or $\phi_1 = \pi - \phi_2$. 
If $\phi_1=\phi_2$, then \eqref{eq:balance2} gives
$
\nd{S} \omega
= 0,
$
which contradicts $\nd{S}>0$ and $\omega>0$. Hence $\phi_1 \neq \phi_2$, and the only remaining solution of \eqref{eq:balance3} is
$
\phi_1 = \pi - \phi_2
$---consistent with the fact that during crawling the segment speeds cannot be in-phase.
 We use these in the remaining harmonic balance equations, obtaining
\begin{small}
\begin{subequations}
\begin{align}
    \frac{4 \pi_\nd{f}}{\pi} (1+ \Delta) \sqrt{1-\nd{a}^2} \cos(\phi_2) & = -2 \zeta \nd{S} \omega + \frac{8 \pi_\nd{V}}{\pi} \frac{\nd{M} \beta}{\nd{S}}, 
        \label{eq:balance5} \\
        \nd{S} (1-\omega^2) & = - \frac{8 \pi_\nd{V} \nd{M}}{\pi} \sqrt{1 - \Big( \frac{\beta}{\nd{S}}\Big)^2}.  
        \label{eq:balance6} 
\end{align}
\label{eq:balance_strain_dyn}
\end{subequations}
\end{small}
 The final set of harmonic balance equations is provided by the dynamics of the center of mass speed \eqref{eq:CoM_dynamics}:
 \begin{small}
\begin{subequations}
    \begin{align}
    \arccos(\nd{a})& = \frac{\pi \Delta}{1+\Delta},
        \label{eq:balance7}\\
        \sin(\phi_2) & = 0. 
        \label{eq:balance8}
    \end{align}
    \label{eq:balance_COM}
\end{subequations}
 \end{small}
\eqref{eq:balance7} yields the value of the unknown $\nd{a}$, revealing that it is only dependent on the parameter $\Delta$ of the frictional model (i.e., friction anisotropy). \eqref{eq:balance8} $\Rightarrow$ $\phi_2 = 0$ or $\phi_2 = \pi$. Since by Assumption \ref{assumption:positive_amplitudes} $\tilde{\nd{v}}>0$ and by \eqref{eq:balance2} $\tilde{\nd{v}} = \frac{\omega \nd{S}}{2 \cos(\phi_2)}$, then $\phi_2 = 0 \; \Rightarrow \, \phi_1 = \pi$. Together, the harmonic balance equalities \eqref{eq:balance_strain_rate}, \eqref{eq:balance_strain_dyn} and \eqref{eq:balance_COM} close the system of equations for the 8 unknowns $\{\omega, \nd{S}, \bar{\nd{v}}_1, \tilde{\nd{v}}_1, \bar{\nd{v}}_2, \tilde{\nd{v}}_2, \phi_1, \phi_2\}$ or alternatively, $\{\omega, \nd{S}, \nd{a}_1, \tilde{\nd{v}}_1, \nd{a}_2, \tilde{\nd{v}}_2, \phi_1, \phi_2\}$. The values of $\nd{a}_1, \nd{a}_2, \phi_1, \phi_2$ have already been determined and $\tilde{\nd{v}}_1 = \tilde{\nd{v}}_2=: \tilde{\nd{v}}$. Thus,  
$\tilde{\nd{v}}, \omega, \nd{S}$ are left to be determined. The respective values are given by \eqref{eq:balance2}, \eqref{eq:balance5} and \eqref{eq:balance6}, which correspond to \eqref{eq:balance_closure} in the main text.

\vspace{-0.5cm}
\bibliographystyle{IEEEtran}
\bibliography{refs}
\vspace{-0.5cm}
\begin{IEEEbiography}[{\includegraphics[width=1in,height=1.25in,clip,keepaspectratio]{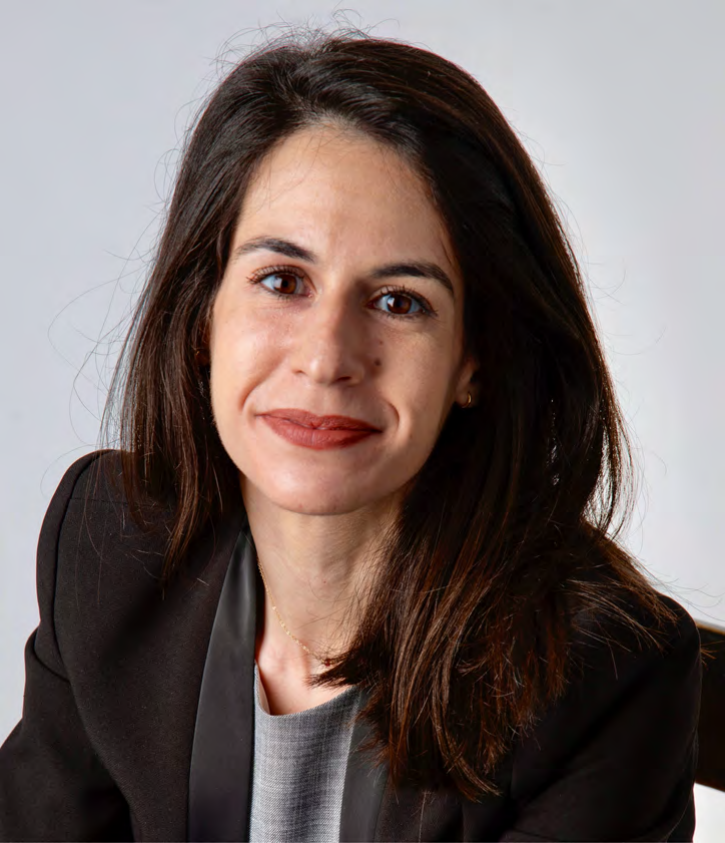}}]{Juncal Arbelaiz} received the B.Sc. and M.Sc. degrees in engineering from the University of Navarre, Spain, and the Ph.D. degree in applied mathematics from the Massachusetts Institute of Technology (MIT) in 2022. She is currently a C.V. Starr Postdoctoral Fellow and Senior Schmidt Science Fellow at Princeton University, affiliated with the Center for Statistics and Machine Learning and the Princeton Neuroscience Institute. Her research interests include optimal decentralized control, optimal estimation, and active learning of spatially distributed dynamical systems. She was named a Rising Star in EECS in 2021 and has received fellowships from the Hugh Hampton Young Memorial Fund, Rafael del Pino Foundation, Google Women Techmakers, la Caixa Foundation, and MIT.
\end{IEEEbiography}
\vspace{-0.5cm}
\begin{IEEEbiography}[{\includegraphics[width=1in,height=1.25in,clip,keepaspectratio]{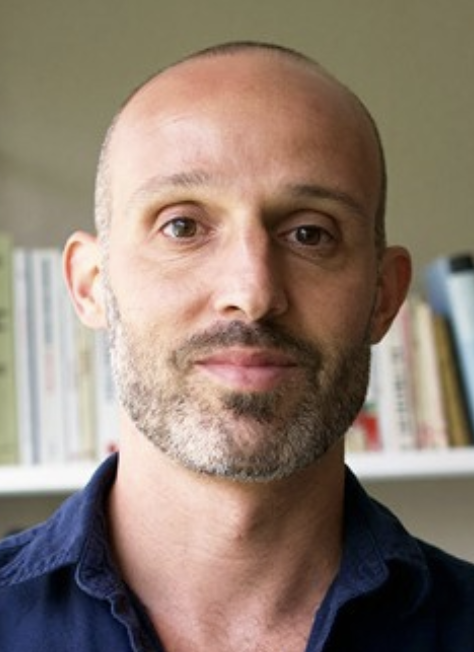}}]{Alessio Franci}
     is a professor in the Department of Electrical Engineering and Computer Science of the University of Liege and co-founder of the ULiege Neuroengineering Lab. He is also an awardee of a WEL-T starting grant, WEL Research Institute. Alessio received his M.Sc. in Theoretical Physics from the University of Pisa in 2008 and his PhD in Physics and Control Theory from the University of Paris Sud 11 in 2012. Between 2012 and 2015 he was a postdoctoral researcher at the University of Liege and INRIA Lille, and a long term visiting researcher at the University of Cambridge. Between 2015 and 2022 he was a professor in the Math Department of the National Autonomous University of Mexico. His research interests are in brain-inspired computing and control theory, neuromorphic engineering, computational neuroscience, and applications to robotics and intelligent sensors.
\end{IEEEbiography}
\vspace{-0.5cm}
\begin{IEEEbiography}[{\includegraphics[width=1in,height=1.25in,clip,keepaspectratio]{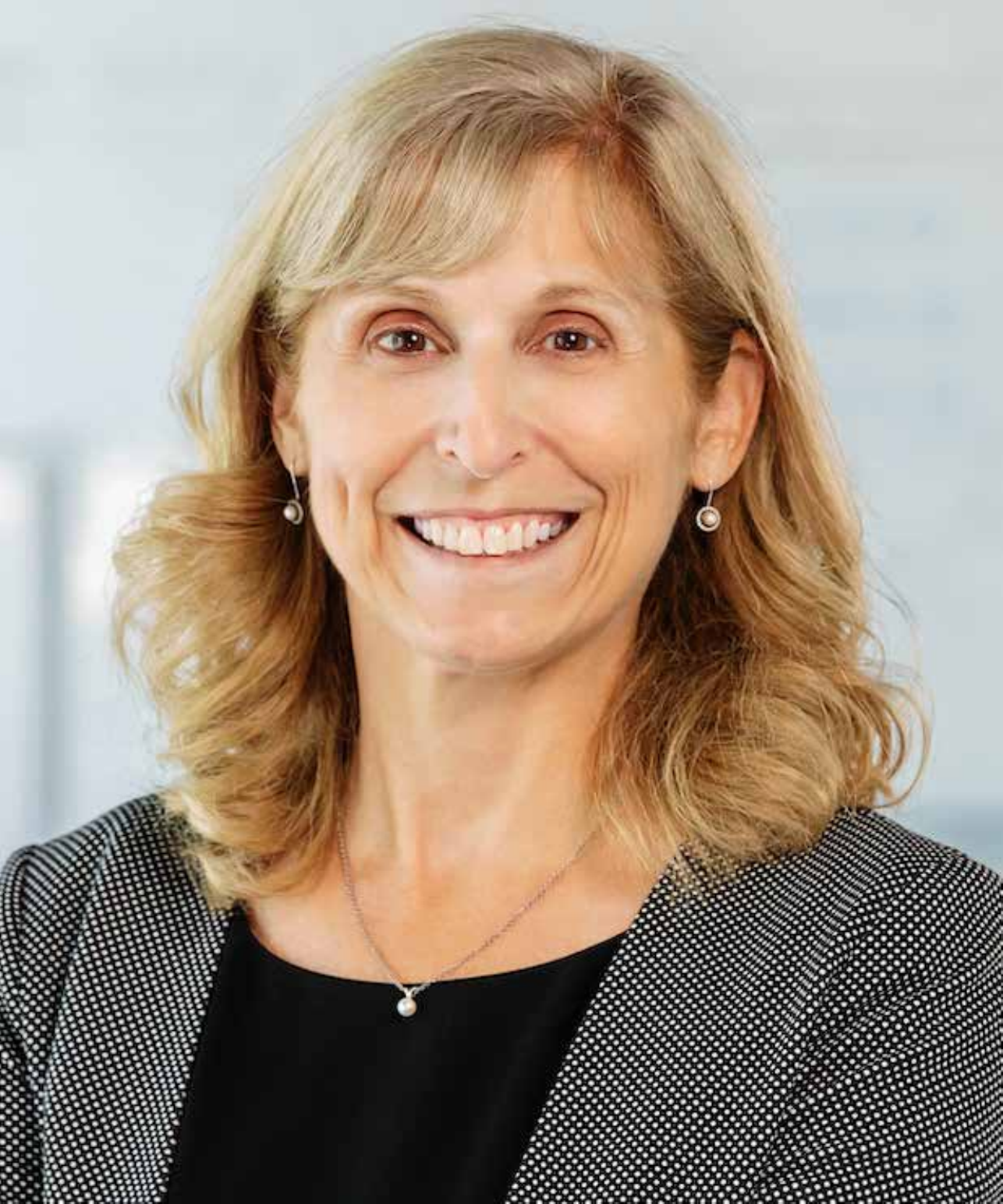}}]{Naomi Ehrich Leonard}
    (F’07) received the B.S.E. degree in mechanical engineering from Princeton University, Princeton, NJ, USA, in 1985, and the M.S. and Ph.D. degrees in electrical engineering from the University of Maryland, College Park, MD, USA, in 1991 and 1994, respectively. 
From 1985 to 1989, she was an Engineer in the electric power industry. She joined Princeton University in 1994 and is currently the Chair and Edwin S. Wilsey Professor with the Department of Mechanical and Aerospace Engineering, an Associated Faculty of the Program in Applied and Computational Mathematics and the Graduate Program in Biophysics, and an Affiliated Faculty with the Princeton Neuroscience Institute. Her research interests include control theory, dynamical systems, networked multiagent systems, robotics, collective animal behavior, and social decision-making. Recent awards include the 2023 IEEE Control Systems Award and the 2024 Richard E. Bellman Control Heritage Award from the American Automatic Control Council. She is a MacArthur Fellow, Fellow of the ASME, IFAC, and SIAM, and an elected member of the American Academy of Arts and Sciences. 
\end{IEEEbiography}
\vspace{-0.5cm}
\begin{IEEEbiography}[{\includegraphics[width=1in,height=1.25in,clip,keepaspectratio]{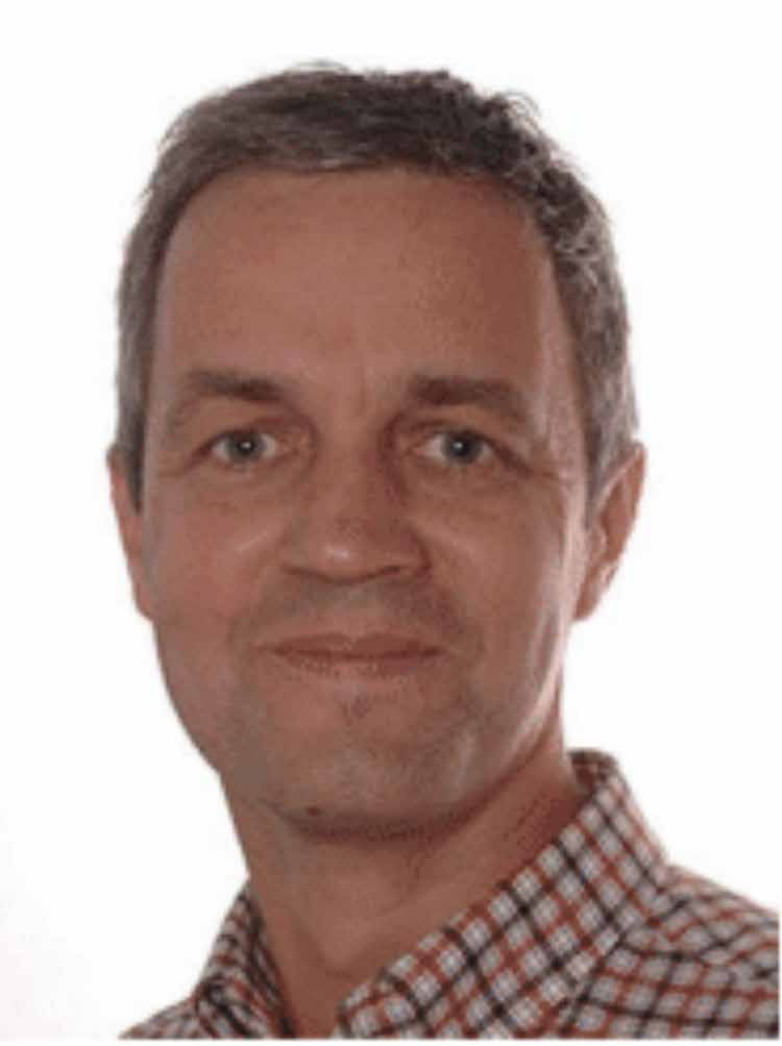}}]{Rodolphe Sepulchre}
   (Fellow, IEEE) received
the engineering degree in in mathematical en-
gineering and the Ph.D. degree in control the-
ory from the Universitat Catholique de Louvain,
Ottignies-Louvain-la-Neuve, Belgium, in 1990
and in 1994, respectively.
He is a Professor of engineering with the University of Cambridge, Cambridge, U.K., since
2013, and at KU Leuven since 2023. His research interests include nonlinear control and
optimization, and more recently neuromorphic
control. He co-authored the monographs “Constructive Nonlinear Control” (Springer-Verlag, 1997) and “Optimization on Matrix Manifolds”
(Princeton University Press, 2008). 
Dr. Sepulchre is a recipient of the IEEE CSS Antonio Ruberti Young
Researcher Prize (2008) and of the IEEE CSS George S. Axelby Outstanding Paper Award (2020). He is a fellow of IEEE, IFAC, and SIAM.
He has been IEEE CSS Distinguished Lecturer between 2010 and 2015.
In 2013, he was elected at the Royal Academy of Belgium.
\end{IEEEbiography}
\vspace{-0.5cm}
\begin{IEEEbiography}[{\includegraphics[width=1in,height=1.25in,clip,keepaspectratio]{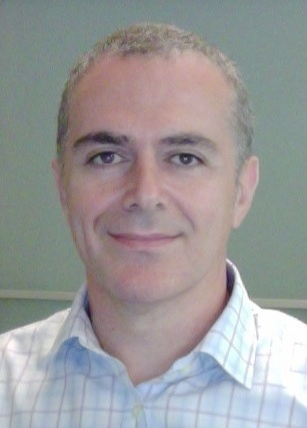}}]{Bassam Bamieh} received the B.Sc. degree in electrical engineering and physics from Valparaiso University, Valparaiso, IN, USA, in 1983, and the M.Sc. and Ph.D. degrees in electrical and computer engineering from Rice University, Houston, TX, USA, in 1986 and 1992, respectively. From 1991 to 1998, he was an Assistant Professor with the Department of Electrical and Computer Engineering, and the Coordinated Science Laboratory, University of Illinois at Urbana-Champaign, Champaign, IL, USA, after which he joined the University of California at Santa Barbara (UCSB), Santa Barbara, CA, USA, where he is currently a Professor of Mechanical Engineering. His research interests include robust and optimal control, distributed and networked control and dynamical systems, shear flow transition and turbulence, and the use of feedback in thermoacoustic energy conversion devices. Dr. Bamieh is a past recipient of the IEEE Control Systems Society G. S. Axelby Outstanding Paper Award (twice), the AACC Hugo Schuck Best Paper Award, and the National Science Foundation CAREER Award. He was elected as a Distinguished Lecturer of the IEEE Control Systems Society in 2005, and a Fellow of the International Federation of Automatic Control (IFAC).
\end{IEEEbiography}

\vfill
\end{document}